\documentclass[pre,aps,twocolumn,superscriptaddress,showpacs]{revtex4-1}%{revtex4}

\usepackage{amssymb}
\usepackage{epsfig}
\usepackage{amsmath}
\usepackage{subfigure}
\usepackage{graphicx}
\usepackage{textcomp}
\usepackage{url}
\usepackage{float}
\usepackage{color}
\usepackage{cases}
\usepackage[colorlinks=true, urlcolor=blue, anchorcolor=blue, citecolor=blue,filecolor=blue,linkcolor=blue,menucolor=blue%,pagecolor=blue
]{hyperref}

\usepackage{color}
\usepackage[normalem]{ulem}

\usepackage{epsfig}
\usepackage{textcomp}
\usepackage{epstopdf}

\newcommand{\antiquad}{\!\!\!\!\!\!\!\!}

\begin{document}
\title{Landau theory of the short-time dynamical phase transitions of the Kardar-Parisi-Zhang
interface}

\author{Naftali R. Smith}
\email{naftali.smith@mail.huji.ac.il}
\affiliation{Racah Institute of Physics, Hebrew University of
Jerusalem, Jerusalem 91904, Israel}

\author{Alex Kamenev}
\email{kamenev@physics.umn.edu}
\affiliation{Department of Physics, University of Minnesota, Minneapolis, MN 55455, USA}
\affiliation{William I. Fine Theoretical Physics Institute, University of Minnesota,
Minneapolis, MN 55455, USA}

\author{Baruch Meerson}
\email{meerson@mail.huji.ac.il}
\affiliation{Racah Institute of Physics, Hebrew University of
Jerusalem, Jerusalem 91904, Israel}


\begin{abstract}

We study the short-time distribution $\mathcal{P}\left(H,L,t\right)$ of the two-point two-time height difference $H=h(L,t)-h(0,0)$ of a stationary Kardar-Parisi-Zhang (KPZ) interface in 1+1 dimension.
Employing the optimal-fluctuation method, we develop an effective Landau theory for the second-order dynamical phase transition found previously for $L=0$ at a critical value $H=H_c$. We show that  $|H|$ and $L$ play the roles of inverse temperature and external magnetic field, respectively.
In particular, we find a first-order dynamical phase transition when $L$ changes sign, at supercritical $H$.
We also determine analytically  $\mathcal{P}\left(H,L,t\right)$ in several limits away from the second-order transition.
Typical fluctuations of $H$ are Gaussian, but the distribution tails are highly asymmetric. The tails $-\ln\mathcal{P}\sim\left|H\right|^{3/2} \! /\sqrt{t}$ and $-\ln\mathcal{P}\sim\left|H\right|^{5/2} \! /\sqrt{t}$, previously found for $L=0$, are enhanced
for $L \ne 0$. At very large $|L|$ the whole height-difference distribution $\mathcal{P}\left(H,L,t\right)$ is time-independent and Gaussian in $H$, $-\ln\mathcal{P}\sim\left|H\right|^{2} \! /|L|$, describing the probability of creating a ramp-like
height profile at $t=0$.

\end{abstract}

\maketitle

\section{Introduction}

The Kardar-Parisi-Zhang (KPZ) equation \citep{KPZ} describes an important universality class of non-equilibrium stochastic growth \citep{HHZ,Barabasi,Krug,QS,S2016,Takeuchi2018}.
In $1+1$ dimension, the KPZ equation reads
\begin{equation}
\label{eq:KPZ_dimensional}
\partial_{t}h=\nu\partial_{x}^{2}h+\frac{\lambda}{2}\left(\partial_{x}h\right)^{2}+\sqrt{D}\,\xi(x,t),
\end{equation}
where $h(x,t)$ is the interface height at the point $x$ of a substrate at time $t$, and $\xi(x,t)$ is a Gaussian noise with zero average
and
\begin{equation}\label{correlator}
\langle\xi(x_{1},t_{1})\xi(x_{2},t_{2})\rangle = \delta(x_{1}-x_{2})\delta(t_{1}-t_{2}).
\end{equation}
The hallmark of the KPZ interface in $1+1$ dimensions is its late-time kinetic roughening scaling properties. The lateral correlation length grows as $t^{2/3}$, and the interface width grows as $t^{1/3}$.

In recent years, more detailed characterizations of the height fluctuations of the KPZ interface have been explored. One of them is the full probability distribution $\mathcal{P}\left(H,t\right)$ of the single-point, two-time interface height difference $H=h\left(x=0,t\right)-h\left(x=0,0\right)$. This distribution depends on the initial condition $h\left(x,t=0\right)$. Remarkably, exact representations for the moment generating function of $\exp[(\lambda/2\nu)H]$, have been obtained for several initial conditions,  see Refs. \citep{Corwin, QS, HHZ} for recent reviews.

This work studies \emph{large deviations} of the KPZ interface height, as manifested by the tails of $\mathcal{P}\left(H,t\right)$.  For some initial conditions the long- and short-time asymptotics of these tails have been extracted from exact representations \citep{DMS, DMRS, SMP, LeDoussal2017}. Such calculations are technically difficult and, more importantly, are limited to the very few cases where exact representations are known.

The optimal fluctuation method (OFM) provides a viable alternative to the exact representations. This approximate method (also known as weak-noise theory, instanton method, macroscopic fluctuation theory, \textit{etc}) originated in condensed matter physics \citep{Halperin,Langer,Lifshitz,Lifshitz1988}. Closely related methods appeared in the studies of turbulence and turbulent transport \citep{turb1,turb2,turb3}, diffusive lattice gases \citep{bertini2015} and
stochastic reactions on lattices \citep{EK, MS2011}. The OFM has already been applied to the KPZ equation and closely related systems in many works \citep{Mikhailov1991, GurarieMigdal1996,Fogedby1998, Fogedby1999,Nakao2003,KK2007,KK2008,KK2009,Fogedby2009,MKV,KMSparabola,Janas2016,MeersonSchmidt2017,
Smith2018,MSV_3d,SmithMeerson2018}.
The method involves a saddle-point evaluation of the path integral of the stochastic process conditioned on a specified large deviation. The minimization procedure generates a classical field theory which can be cast into Hamiltonian form. The solution of the Hamilton equations yields
the optimal (most likely) path of the system and the most likely realization of the noise. With the solution at hand, $-\ln\mathcal{P}$ can be found (up to a pre-exponential factor) by evaluating the ``classical'' action along the optimal path.

In this work we focus on the stationary initial condition, where it is assumed that the interface has evolved for an infinitely long time prior to $t=0$. A statistical ensemble of initial interface configurations  $h\left(x,t=0\right)$ is given by  random realizations of a two-sided Brownian motion:
\begin{equation}
\label{incond1}
h\left(x,t=0\right) = \frac{\nu}{D} \, B(x),
\end{equation}
where $B(x)$ is the two-sided Wiener process with diffusion constant $1$ \citep{pinned}. For this initial condition Imamura and Sasamoto \citep{IS} and Borodin \textit{et al.} \citep{Borodinetal} obtained exact representations for $\mathcal{P}\left(H,t\right)$ in terms
of Fredholm determinants. They also proved that, in the long-time limit, $t\gg \nu^{5}/(D^{2}\lambda^{4})$, and in a proper moving frame \citep{footnote:displacement},
the typical fluctuations of the single-point height difference scale with time as $t^{1/3}$, in agreement with the exponent $1/3$ of the interface width growth,
and that the distribution $\mathcal{P}$ of the typical fluctuations is the Baik-Rains distribution \citep{BR}.

The \emph{short}-time  behavior, $t\ll \nu^{5}/(D^{2}\lambda^{4})$, of $\mathcal{P}\left(H,t\right)$ was studied by Janas \textit{et al.}~\citep{Janas2016}. Using the OFM, they found that %, in a proper moving frame \citep{footnote:displacement},
the short-time scaling form of the height distribution is $-\ln\mathcal{P}\left(H,t\right)\simeq s\left(H\right)/\sqrt{t}$. Janas \textit{et al.} \citep{Janas2016} calculated the large-deviation function $s(H)$ analytically in several limits and also computed it numerically.
They found that the short-time $\lambda H \to +\infty$ tail, $-\ln\mathcal{P}\left(H,t\right)\sim\left|H\right|^{3/2} \! /\sqrt{t}$, coincides with that of the Baik-Rains distribution, and conjectured that this tail is valid at all times $t>0$. They also conjectured that the $\lambda H \to -\infty$ tail, $-\ln\mathcal{P}\left(H,t\right)\sim\left|H\right|^{5/2} \! /\sqrt{t}$, persists at long times for $|H|\gg t$. A similar conjecture \cite{KMSparabola} for the ``droplet'' initial condition was recently proven to be correct \citep{SMP,CorwinGhosal,KLD2018,Corwinetal2018}.

Importantly, Janas \textit{et al.} \cite{Janas2016} uncovered a singularity -- a jump in the second derivative -- of the large-deviation function $s(H)$ with respect to $H$ at a critical value of $\lambda H=\lambda H_c>0$. As they showed, this singularity is caused by a spontaneous breaking  of the spatial reflection symmetry $x \leftrightarrow -x$ of the optimal path of the  interface. Subsequently, Krajenbrink and Le Doussal \citep{LeDoussal2017} determined the whole large deviation function $s(H)$ exactly, and reproduced the singularity at $H=H_c$, by extracting the short-time asymptotics from the exact representation
\citep{IS,Borodinetal} for $\mathcal{P}\left(H,t\right)$.

Large-deviation functions of nonequilibrium systems can be viewed as analogs of equilibrium free energy. Therefore, it is natural to interpret their singularities
as (dynamical) phase transitions (DPTs). Such transitions, of the first and second order, have been found in several non-equilibrium models of lattice gases, see Refs. \cite{Schuetz,Derrida2007,hurtadoreview,bertini2015} for reviews.
It is appealing to characterize these systems in terms of (a nonequilibrium extension of) Landau theory of phase transitions \cite{Stanley}, and this has been already done for some of these models \citep{Baek2015,Baek2017,Baek2018}.

In this paper we extend the short-time analysis of Refs. \citep{Janas2016,LeDoussal2017} in two directions.
First, we develop an effective Landau theory of the second-order short-time dynamical phase transition at $H=H_c$.
We introduce an order parameter which quantifies the spatial-reflection asymmetry of the optimal path of the interface.
As a result, the large deviation function of the height as a function of the order parameter plays a role similar to that of the equilibrium free energy in the standard Landau theory \citep{Stanley}. Second, we generalize the problem by studying the probability distribution $\mathcal{P}(H,L,t)$ of the \emph{two-point} height difference $H=h(L,t)-h(0,0)$. An exact representation for the distribution of this quantity  is unknown.   We find that, in the vicinity of the critical point $H=H_c$, the quantities $\lambda H$ and $L/\sqrt{t}$ play the roles of inverse temperature and external magnetic field, respectively, of the equilibrium second-order phase transition.
Our effective Landau theory yields a detailed characterization of the dynamical phase transition in terms of the critical exponents which describe the singular behaviors of the order parameter and of the large-deviation function of $\mathcal{P}(H,L,t)$ as one approaches $L=0$ and $H=H_c$. In particular, we find that, at supercritical $H$, a change in sign of $L$ is accompanied by a first-order dynamical phase transition.

Away from the second-order phase transition, we determine the scaling forms of $\mathcal{P}\left(H,L,t\right)$, and calculate the corresponding scaling functions and coefficients, in the following limits.
For arbitrary $H$ and sufficiently large $L/\sqrt{t}$, the height-difference distribution is Gaussian and independent of time:
\begin{equation}
\label{eq:stationary_ramp_P}
-\ln\mathcal{P}\left(H,L,t\right)\simeq\frac{\nu H^{2}}{D\left|L\right|}.
\end{equation}
For small $H$, the process is approximately described by the Edwards-Wilkinson (EW) equation \citep{EW1982}, and the height-difference distribution is Gaussian but, in general, time-dependent:
\begin{equation}\label{gauss1}
-\ln\mathcal{P}\left(H,L,t\right)\simeq\frac{\nu^{1/2}H^{2}}{D\sqrt{t}}\,g\left(\frac{L}{\sqrt{\nu t}}\right).
\end{equation}
The scaling function $g(\dots)$ is described by Eq.~(\ref{eq:scaling_function_EW}) below. It decreases monotonically as a function of $|\ell|=|L|/\sqrt{\nu t}$. At $\ell =0$ $g=\sqrt{\pi}/2$ in agreement with previous work \citep{Janas2016,Krug1992}. At large $|\ell|$
$g(\ell)\simeq 1/|\ell|$, and Eq.~(\ref{gauss1}) %$g\left(\left|\ell\right|\to\infty\right)\simeq1/\left|\ell\right|$, in agreement with
coincides with Eq.~(\ref{eq:stationary_ramp_P}).

The tails of the height-difference distribution are non-Gaussian and asymmetric. For large positive $\lambda H$ we find the following scaling behavior:
\begin{equation}
\label{s_traveling_wave}
-\ln\mathcal{P}\left(H,L,t\right)\simeq
\frac{\nu\left|H\right|^{3/2}}{D\sqrt{\left|\lambda\right|t}}\,f\left(\frac{L}{\sqrt{\lambda Ht}}\right)
\end{equation}
The scaling function $f(\dots)$ is given by Eq.~(\ref{eq:scaling_function_negative_tail}) below. It decreases monotonically as a function of $|\eta|$, where $\eta = L/\sqrt{\lambda Ht}$. At $L=0$ we obtain $f=4\sqrt{2}/3$, which corresponds to the $\left|H\right|^{3/2} \! /\sqrt{t}$ tail of the Baik-Rains distribution \citep{BR,Janas2016,LeDoussal2017}. The large-$|\eta|$ asymptote, $f\left(\left|\eta\right|\gg1\right)\simeq1/\left|\eta\right|$, is consistent with Eq.~(\ref{eq:stationary_ramp_P}).

Finally, for large negative $\lambda H$ we obtain
\begin{equation}
\label{eq:action_HD_tail}
-\ln\mathcal{P}\left(H,L,t\right)\simeq\frac{4\sqrt{2}}{15\pi D\lambda^{2}\sqrt{t}}\left(-\lambda H-\frac{L^{2}}{2t}\right)^{5/2}.
\end{equation}
This tail is independent of $\nu$. For $L=0$ it reproduces the $\left|H\right|^{5/2} \! /\sqrt{t}$ tail found previously \citep{Janas2016,LeDoussal2017}.

All the asymptotic results (\ref{eq:stationary_ramp_P})-(\ref{eq:action_HD_tail}) show that the probability of observing an unusually large $|H|$ for stationary interface increases with $|L|$. This important observation is also supported by our numerics for moderate $H$, not captured by these asymptotics.

The remainder of this paper is organized as follows. In Sec. \ref{sec:OFM} we present the OFM formulation of the problem.
In Sec. \ref{sec:DPT} we define a proper order parameter and develop the effective Landau theory: first for $L=0$ and then for $L\ne 0$.
In Sec. \ref{sec:OFMsol} we obtain the asymptotics (\ref{eq:stationary_ramp_P})-(\ref{eq:action_HD_tail}) of the height-difference distribution,
and the corresponding optimal paths of the interface.
We summarize and discuss our results in Sec. \ref{disc}.

\section{Optimal fluctuation method}

\label{sec:OFM}

\subsection{OFM equations and constraints}

As we already mentioned, the OFM has been employed for the analysis of the KPZ equation in many papers \citep{Mikhailov1991,GurarieMigdal1996,Fogedby1998, Fogedby1999,Nakao2003, KK2007,KK2008,KK2009,Fogedby2009,MKV,KMSparabola,Janas2016,MeersonSchmidt2017,Smith2018,MSV_3d,SmithMeerson2018}. For the two-sided Brownian interface (\ref{incond1}), the derivation of the governing equations closely follows that of  Ref. \citep{Janas2016}, so we can be brief.

We introduce the observation time $T$ at which the interface height difference, $h\left(L,T\right)-h\left(0,0\right)=H $, is measured.
We assume, without loss of generality, that $\lambda<0$ \citep{footnote:signlambda}.
The rescaling $\tilde{x} = x/\sqrt{\nu T}$, $\tilde{t} = t/T$, $\tilde{h} = \left|\lambda\right|h/\nu$ brings
Eq.~(\ref{eq:KPZ_dimensional}) to the dimensionless form \citep{MKV}
\begin{equation}
\label{eq:KPZ_dimensionless}
\partial_{t}h=\partial_{x}^{2}h-\frac{1}{2}\left(\partial_{x}h\right)^{2}+\sqrt{\epsilon} \, \xi\left(x,t\right),
\end{equation}
where $\epsilon=D\lambda^{2}\sqrt{T}/\nu^{5/2}$ is the rescaled noise magnitude, and we suppress the tildes for brevity.
The interface height difference $H$ (rescaled by $\nu / |\lambda|$) is measured between the (rescaled) points  $x=0$ and $x = \ell = L/\sqrt{\nu T}$, that is,
\begin{equation*}
H=h\left(\ell,1\right)-h\left(0,0\right).
\end{equation*}

In the weak-noise (that is, short-time) limit, $\epsilon \ll 1$,
one can evaluate the proper path integral of Eq.~(\ref{eq:KPZ_dimensionless}) via the saddle-point method.
This leads to a minimization problem for the effective action. For the stationary interface, the effective action has two terms: $s=s_{\text{dyn}} + s_{\text{in}}$, where
\begin{equation}
\label{eq:sdyn_def}
s_{\text{dyn}}=\frac{1}{2}\int_{0}^{1}dt\int_{-\infty}^{\infty}dx\left[\partial_{t}h-\partial_{x}^{2}h+\frac{1}{2}\left(\partial_{x}h\right)^{2}\right]^{2}
\end{equation}
is the dynamic contribution, and
\begin{equation}
\label{cost}
s_{\text{in}}=\int_{-\infty}^{\infty}dx\left.\left(\partial_{x}h\right)^{2}\right|_{t=0}
\end{equation}
is the ``cost'' of the initial height profile \citep{Janas2016}.
It is convenient to recast the ensuing Euler-Lagrange equation into two Hamiltonian equations for two canonically conjugated fields: $h\left(x,t\right)$ -- the optimal history of the height profile, and $\rho\left(x,t\right)$ -- the optimal realization of the noise $\xi$. The Hamiltonian equations are \citep{Fogedby1998, KK2007, MKV}
\begin{eqnarray}
  \partial_{t} h &=& \frac{\delta\mathcal{H}}{\delta\rho} = \partial_{x}^2 h - \frac{1}{2} \left(\partial_x h\right)^2+\rho ,  \label{eqh}\\
  \partial_{t}\rho &=& - \frac{\delta\mathcal{H}}{\delta h} = - \partial_{x}^2 \rho - \partial_x \left(\rho \partial_x h\right) ,\label{eqrho}
\end{eqnarray}
where
\begin{equation*}
\mathcal{H}=\int_{-\infty}^{\infty}\!dx\,\rho\left[\partial_{x}^{2}h-(1/2)\left(\partial_{x}h\right)^{2}+\rho/2\right]
\end{equation*}
is the Hamiltonian.
Note that $\rho$ undergoes rescaling  $|\lambda| T \rho/\nu \to \rho$.
The condition $h\left(x=\ell,t=1\right)=H$ leads to
\begin{equation}
\label{pT}
\rho\left(x,1\right)=\Lambda_{1}\,\delta\left(x-\ell\right),
\end{equation}
where $\Lambda_{1}$ is a Lagrange multiplier, ultimately determined by the rescaled $H$.
The initial condition for the stationary interface follows from the variation of the total action functional $s$ over $h(x,t=0)$ and takes the form \citep{Janas2016}
\begin{equation}
\label{eq:OFM_initial_condition}
\rho\left(x,t=0\right)+2\partial_{x}^{2}h\left(x,t=0\right)=\Lambda_{1}\delta\left(x\right).
\end{equation}
To prevent the action from diverging, $\rho\left(x,t\right)$ and $\partial_{x}h\left(x,0\right)$ must decay sufficiently rapidly at $\left|x\right|\to\infty$. Finally, we require
\begin{equation}
\label{eq:BC_0_and_H}
h\left(x=0,t=0\right)=0 \;\; \text{and} \;\; h\left(x=\ell,t=1\right)=H.
\end{equation}
The first equality is simply a convenient choice of the reference frame. After solving the OFM problem, we can evaluate $s=s_{\text{dyn}} + s_{\text{in}}$, where $s_{\text{dyn}}$ can be recast as
\begin{equation}
\label{actiongeneral}
s_{\text{dyn}}=\frac{1}{2}\int_{0}^{1}dt\int_{-\infty}^{\infty}dx\,\rho^{2}\left(x,t\right).
\end{equation}
From here we can obtain $\mathcal{P}$ up to a pre-exponential factor: $-\ln \mathcal{P} \simeq s/\epsilon$, or
\begin{equation}
-\ln\mathcal{P}\left(H,T,L\right)
\simeq\frac{\nu^{5/2}}{D\lambda^{2}\sqrt{T}}\,\,s\left(\frac{\left|\lambda\right|H}{\nu},\frac{L}{\sqrt{\nu T}}\right).
 \label{actiondgen}
\end{equation}
in the physical variables.
The action $s$ is the large deviation function of the height-difference distribution at $T \to 0$.

It has been recently shown  that, in addition to the standard KPZ symmetries \citep{Takeuchi2018}, the KPZ equation in $1+1$ dimension has an additional symmetry  \citep{Frey1996, Canet2011, Mathey2017, SmithMeerson2018}. At the level of the OFM, this symmetry is manifested in the invariance of Eqs.~(\ref{eqh}) and~(\ref{eqrho}) under the transformation
\begin{equation}\label{newsymmetry}
-h \! \left(x,-t\right) \! \to \! h \! \left(x,t\right), \quad \rho\left(x,-t\right)+2\partial_{x}^{2}h \! \left(x,-t\right) \! \to \! \rho\left(x,t\right).
\end{equation}
A remarkable property of the stationary interface is that this additional symmetry is respected by the boundary conditions in time. More precisely,
the entire OFM problem~(\ref{eqh})-(\ref{eq:BC_0_and_H}) is invariant under the transformation
\begin{eqnarray}
\label{eq:transform_h}
&&H-h\left(\ell-x,1-t\right)\to h\left (x,t\right), \\
\label{eq:transform_rho}
&&\rho\left(\ell-x,1-t\right)+2\partial_{x}^{2}h\left(\ell-x,1-t\right) \to \rho\left(x,t\right),
\end{eqnarray}
which involves, in addition to the symmetry (\ref{newsymmetry}), the well-known mirror-reflection KPZ symmetry $x\leftrightarrow -x$. It immediately follows that, in the regime of parameters where there is a unique solution to the OFM problem, the solution must respect the  symmetry (\ref{eq:transform_h}) and (\ref{eq:transform_rho}). In particular, the optimal interface history must obey the combined symmetry
\begin{equation}
\label{eq:nontrivial_symmetry}
h\left(x,t\right)=H-h\left(\ell-x,1-t\right).
\end{equation}
Where multiple solutions to the OFM problem exist, some of them could, in principle, break its symmetries. However,  we found, through perturbative analytical solutions and numerics (see below), that the combined symmetry~(\ref{eq:nontrivial_symmetry}) is respected even when multiple solutions exist. Moreover, we argue that for $\ell=0$, the symmetry~(\ref{eq:nontrivial_symmetry}) must hold, because if it were spontaneously broken, one of the branches of the large-deviation function $s(H)$ would have an additional singularity (besides the singularity at $H=H_c$). However, we know from the exact solution \citep{LeDoussal2017} that this is not the case.
We exploit the symmetry~(\ref{eq:nontrivial_symmetry}) below when we have to
choose the correct solution out of families of solutions of  reduced problems.

\section{Dynamical phase transition}
\label{sec:DPT}

\subsection{$\ell=0$}
\label{sec:PT_ell0}

As found in Refs. \citep{Janas2016, LeDoussal2017}, for $\ell=0$, a second-order dynamical phase transition occurs at $H = H_c =-3.70632489 \dots$. For $H<H_{c}$ (which, in view of $H_c<0$ are supercritical heights) the optimal history $h(x,t)$ spontaneously breaks the spatial reflection symmetry $x\leftrightarrow-x$, causing a non-analyticity of the large deviation function $s\left(H\right)$ at $H=H_c$.
In the subcritical region, $H>H_{c}$, the problem~(\ref{eqh})-(\ref{eq:BC_0_and_H}) admits a unique solution for the optimal path which, at all rescaled times $0\leq t\leq 1$, is symmetric with respect to $x$. In the supercritical region $H<H_{c}$
the problem~(\ref{eqh})-(\ref{eq:BC_0_and_H}) has three solutions: a (non-optimal) spatially-symmetric one and two additional spatially-asymmetric solutions which are mirror reflection of each other around $x=0$ \citep{Janas2016}. This situation calls for
an effective Landau theory which we now formulate.

We start by choosing a suitable order parameter, which quantifies the asymmetry of the optimal interface at $t=0$ \citep{footnote:Delta}. As such we adopt the difference between the initial ``costs'' of the right ($x>0$) and left ($x<0$) halves of the system,
\begin{equation}
\label{eq:Delta_def0}
\delta s_{\text{in}}\equiv\int_{0}^{\infty}dx\left.\left(\partial_{x}h\right)^{2}\right|_{t=0}-\int_{-\infty}^{0}dx\left.\left(\partial_{x}h\right)^{2}\right|_{t=0}.
\end{equation}
Next, we define a nonequilibirum analog of the Landau free energy $F(H,\Delta)$ as the minimum of the total action $s=s_{\text{in}}+s_{\text{dyn}}$ under \emph{two} constraints: $h(0,1)=H$ and
\begin{equation}
\label{eq:Delta_def}
\int_{0}^{\infty}dx\left.\left(\partial_{x}h\right)^{2}\right|_{t=0}
-\int_{-\infty}^{0}dx\left.\left(\partial_{x}h\right)^{2}\right|_{t=0} = \Delta ,
\end{equation}
and the additional condition $h(0,0)=0$. The true action, unconstrained by Eq.~(\ref{eq:Delta_def}), is then obtained via an additional minimization over $\Delta$:
\begin{equation}
s\left(H\right)=\min_{\Delta}F\left(H,\Delta\right).
\end{equation}
As we will see shortly, $-H$ plays the role of inverse temperature of equilibrium systems.
The new constraint~(\ref{eq:Delta_def}) can be incorporated into the minimization procedure of $s$ via an additional Lagrange multiplier $\Lambda_{2}$. This results in a modification of the initial condition for the OFM problem, so that Eq.~(\ref{eq:OFM_initial_condition}) gives way to
\begin{equation}
\label{eq:OFM_initial_condition_with_lambda2}
\rho\!\left(x,t=0\right)+2\partial_{x} \! \left\{ \left[1\!+\!\text{sgn}\left(x\right) \! \Lambda_{2}\right]\partial_{x}h\!\left(x,t=0\right)\right\} =\Lambda_{1}\delta\left(x\right).
\end{equation}
The values of $\Lambda_{1}$ and $\Lambda_{2}$ are ultimately set by $H$ and $\Delta$.
%%%

When $\Lambda_2=0$, Eq.~(\ref{eq:OFM_initial_condition_with_lambda2}) coincides with Eq.~(\ref{eq:OFM_initial_condition}).
Therefore, we expect $\Lambda_2$ to vanish  for all the solutions of the
OFM problem~(\ref{eqh})-(\ref{eq:BC_0_and_H}): for the unique solution at $H>H_c$ and for the three solutions (the non-optimal and the two optimal)
at $H<H_c$. Further, the solutions, for which the second Lagrange multiplier $\Lambda_{2}$ vanishes, should correspond to local extrema of the ``free energy'' $F\left(H,\Delta\right)$ as a function of $\Delta$.
We therefore expect $F\left(H,\Delta\right)$ to have an extremum at $\Delta = 0$ at all $H$ and, in addition, two extrema at $\Delta=\pm\Delta_*\ne0$ for supercritical $H$.
As we now show, our numerical results fully support these predictions.
%%%

\begin{figure}[ht]
\includegraphics[width=0.235\textwidth,clip=]{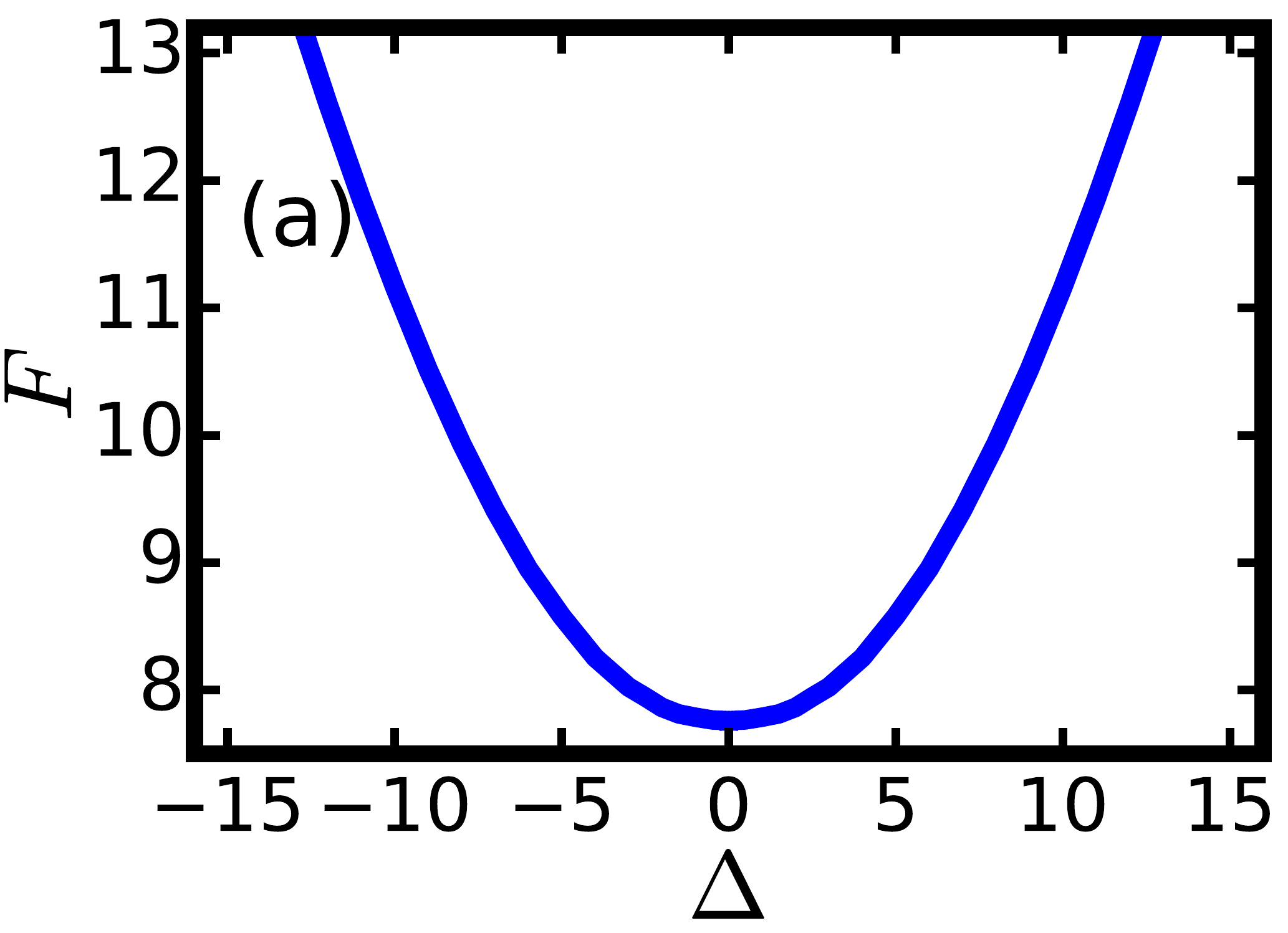}
\includegraphics[width=0.235\textwidth,clip=]{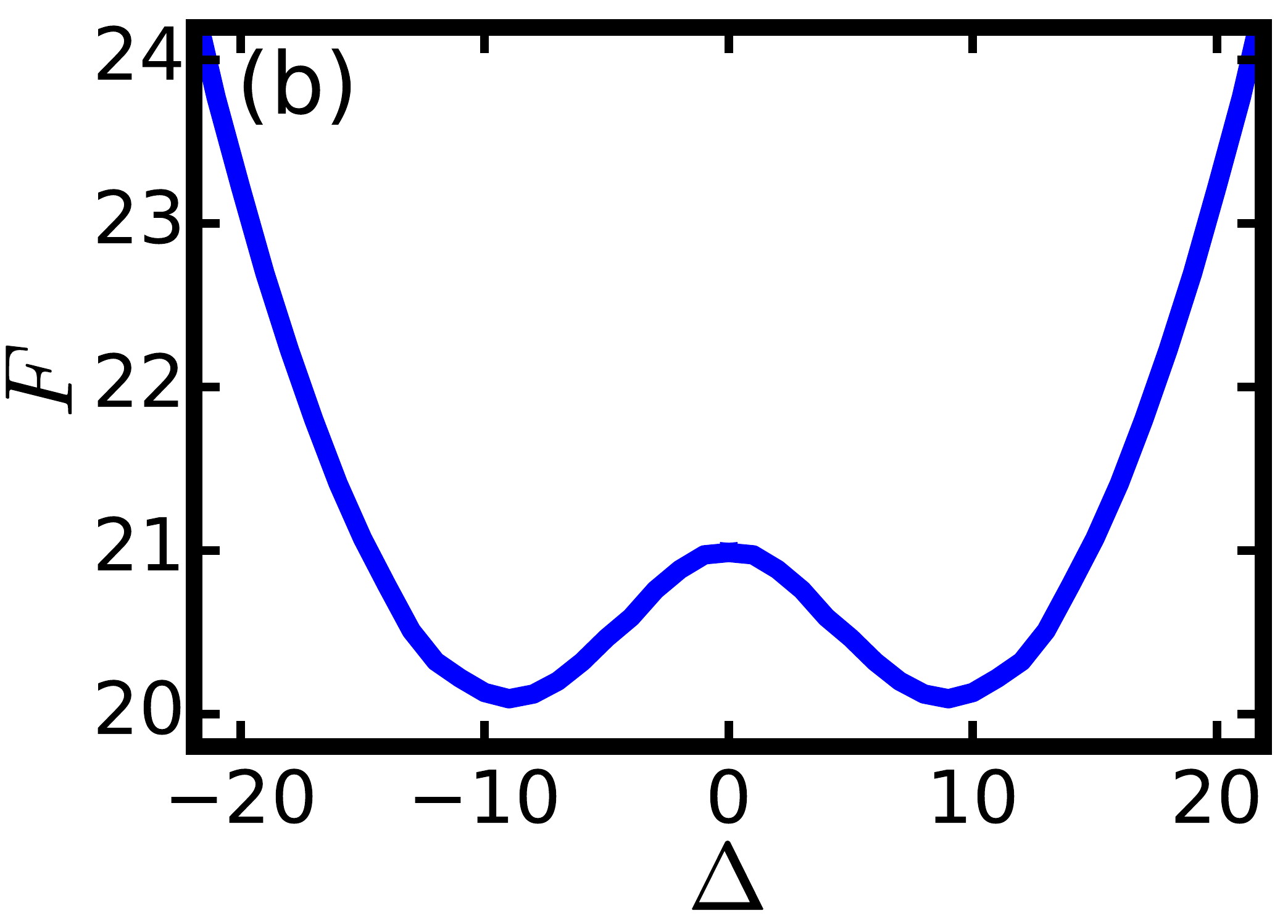}
\includegraphics[width=0.235\textwidth,clip=]{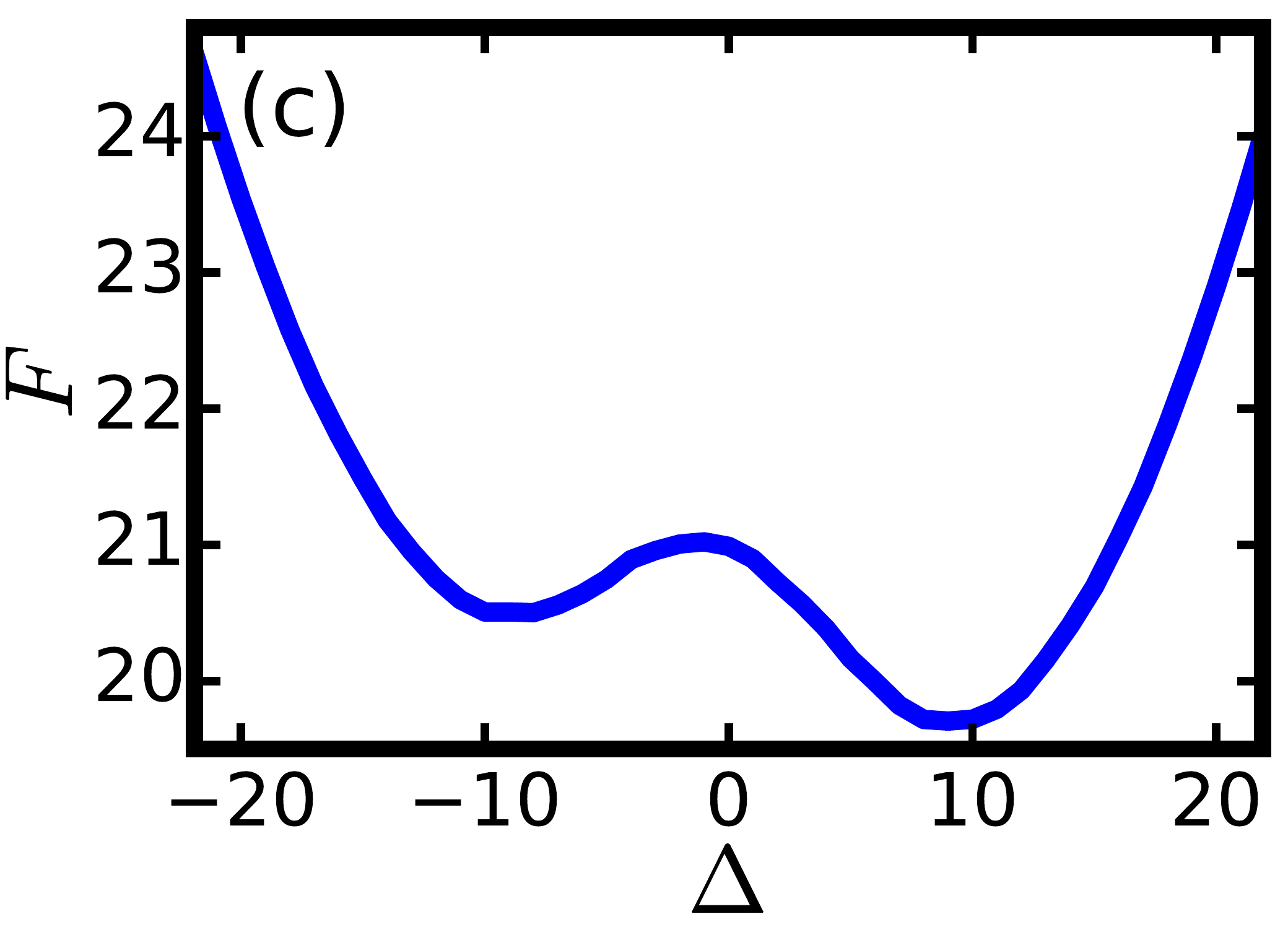}
\includegraphics[width=0.235\textwidth,clip=]{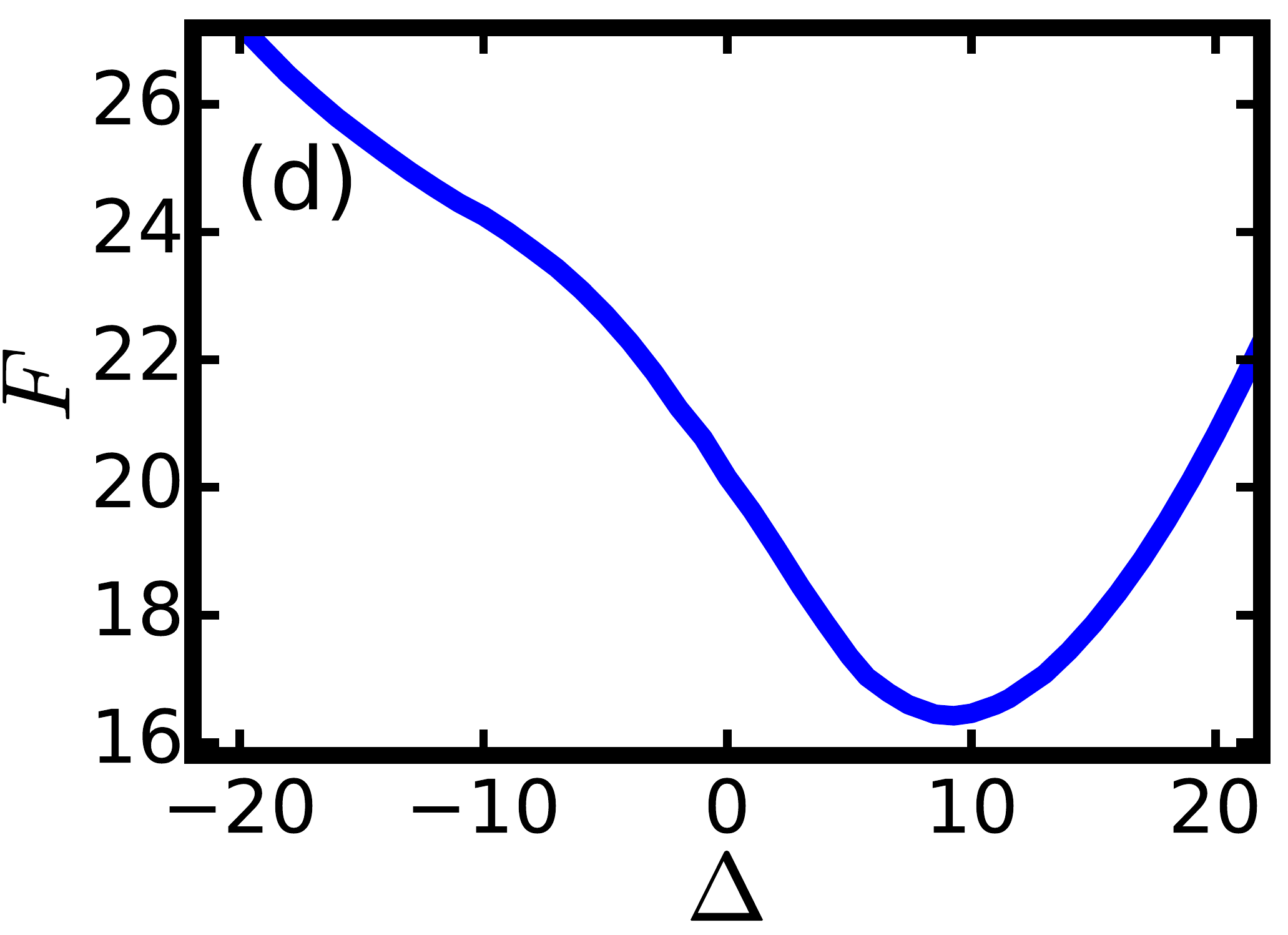}
\caption{The effective Landau free energy $F\left(H,\Delta,\ell\right)$ as a function of $\Delta$ at $\ell=0$, $H=-3$ (a), $\ell=0$, $H=-5$ (b), $\ell=0.05$, $H=-5$ (c) and $\ell=0.5$, $H=-5$ (d).
The parameters $\Delta$, $-H$ and $\ell$ take the roles of order parameter, inverse temperature and external magnetic field, respectively.}
\label{fig:F_L}
\end{figure}

Figures~\ref{fig:F_L} (a) and (b) show $F(H,\Delta)$ as a function of $\Delta$ at fixed $H$, for $H=-3$ and $H=-5$ respectively. We obtained these results by solving the OFM equations with the Chernykh-Stepanov back-and-forth iteration algorithm \citep{Chernykh}. The results strongly support the Landau picture: in the regime $\left|H\right|<\left|H_{c}\right|$, the minimum of $F$ is at $\Delta=0$, whereas in the regime $\left|H\right|>\left|H_{c}\right|$, $\Delta=0$ becomes the point of a local maximum of $F$, and there are two minima at $\Delta=\pm\Delta_*$.
As we verified numerically (not shown), the transition between the two regimes indeed occurs at $H=H_{c}$.
For supercritical $H$ the dependence of $\Delta_*$ on $H$ near the transition is predicted by Landau theory to be $\Delta_{*} \sim \left(H_c-H\right)^{1/2}$ (corresponding to the critical exponent $\beta=1/2$ \citep{Stanley}).
The large deviation function $s$ exhibits a jump in its second derivative, $\partial_{H}^{2}s$, at $H=H_c$ \citep{Janas2016, LeDoussal2017}.
This corresponds to the critical exponent $\alpha = 0$ which describes the behavior $\partial_{H}^{2}s\sim\left|H-H_{c}\right|^{-\alpha}$ of the ``specific heat'' near the phase transition, also in accordance with Landau theory \citep{Stanley}.
%The corresponding critical exponent is $\alpha = 0$, also in accordance with Landau theory \citep{Stanley}.

\subsection{$\ell\ne0$ plays the role of external magnetic field}
\label{sec:landau_theory_nonzero_ell}
We now extend our analysis to $\ell\ne0$ by writing
\begin{equation}
s\left(H,\ell\right)=\min_{\Delta}F\left(H,\Delta,\ell\right)
\end{equation}
and extending the definition of $F$ to nonzero $\ell$ by modifying the constraint at $t=1$ to $h\left(\ell,1\right)=H$.
Near the phase transition, $\left|H-H_{c}\right|\ll\left|H_{c}\right|$, the parameter $\ell$ has a role analogous to the external magnetic field in Landau theory \citep{Stanley}.
Indeed, for $\ell\ne0$ the spatial reflection symmetry $x \leftrightarrow -x$ is broken, and the minimum of $F$ is at a nonzero $\Delta_*$ even for subcritical $H$. For supercritical $H$, a small but nonzero $\ell$ causes one of the minima of $F\left(\Delta\right)$ to be lower than the other, making it optimal. For larger $\ell$, only one minimum remains.
Our numerical solutions demonstrate these features in Figs.~\ref{fig:F_L} (c) and (d) for $H=-5$ and two different nonzero values of $\ell$: $0.05$ and $0.5$.
Overall, Fig.~\ref{fig:F_L} suggests that, in the vicinity of $H=H_c$ and $\ell=\Delta=0$, $F$ has the standard mean-field Landau form
\begin{equation}
\label{eq:F_L_form}
F \! \left(H,\Delta,\ell \right) \!= \! F_{0} \! \left(H\right)+\alpha_{1} \! \left(H-H_{c}\right) \! \Delta^{2} \! +\alpha_{2}\Delta^{4} \! -\alpha_{3}\ell \Delta +\dots
\end{equation}
with $\alpha_{1,2,3}>0$.
This effective Landau theory yields two additional critical exponents. The ``susceptibility'' diverges at the transition  as \citep{Stanley}
\begin{equation}
\left.\frac{\partial\Delta_{*}}{\partial\ell}\right|_{\ell=0}\sim\left|H-H_{c}\right|^{-\gamma}
\end{equation}
with $\gamma=1$, and the dependence of the order parameter on the ``external magnetic field'' $\ell$ at $H=H_c$ is
\begin{equation}
\left.\Delta_{*}\right|_{H=H_{c}}\sim\ell^{1/\delta}
\end{equation}
with $\delta=3$ \citep{Stanley}.
That the critical exponents $\alpha$, $\beta$, $\gamma$ and $\delta$ all take their mean-field Landau theory values follows directly from Eq.~(\ref{eq:F_L_form}).

The effective Landau theory implies that, at fixed supercritical $H$, there is a \emph{first-order} dynamical phase transition, corresponding to a jump of $\partial s/\partial\ell$, when $\ell$ changes sign.
This transition occurs because the optimal path switches between two asymmetric solutions as $\ell$ changes sign. This feature is most easily seen in the $-H \gg 1$ tail, where the optimal history can be found analytically, see Sec.~\ref{negtail} below.

We checked that, for $\ell\neq 0$ and $\Lambda_2=0$, our numerical solutions exhibit the combined symmetry (\ref{eq:nontrivial_symmetry}). We do not show these plots here.

\section{Perturbative solutions}

\label{sec:OFMsol}

\begin{figure}[ht]
\includegraphics[width=0.45\textwidth,clip=]{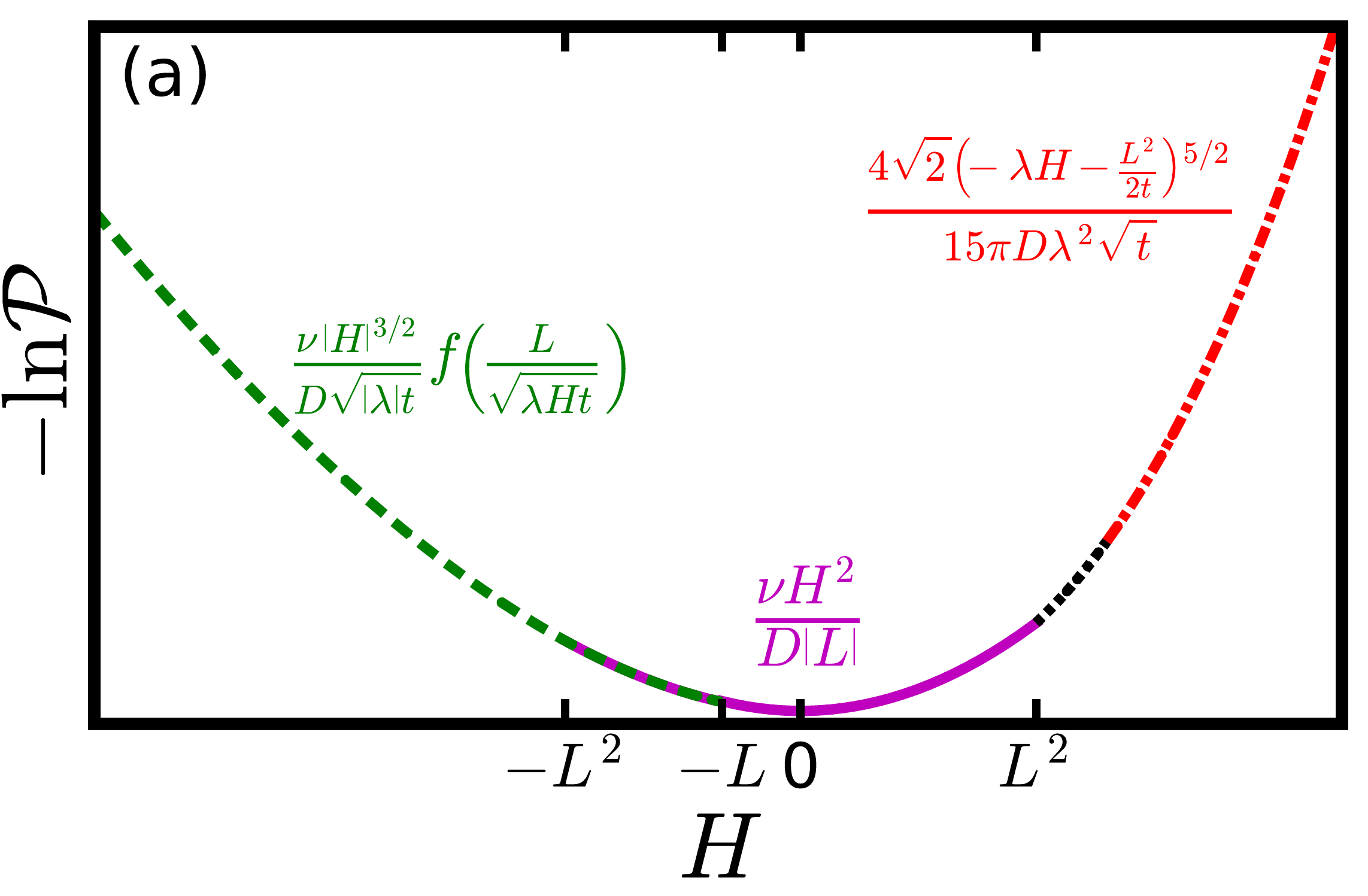}
\includegraphics[width=0.45\textwidth,clip=]{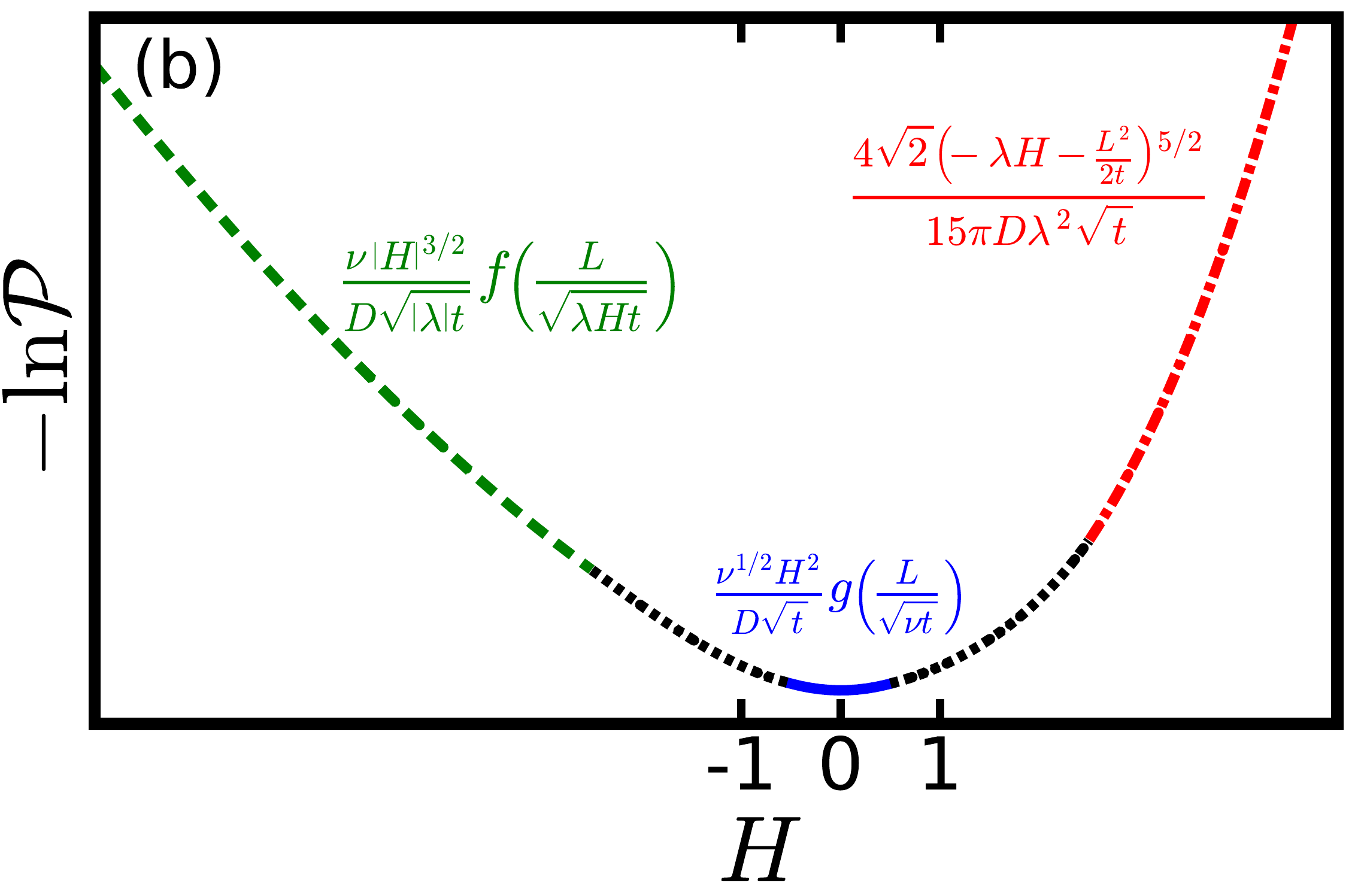}
\caption{The minus logarithm of the short-time ($t \ll \nu^5 / D^2 \lambda^4$) height-difference distribution at fixed $L$ and $t$, plotted schematically, for $L \gg \sqrt{\nu t}$ (a) and $L \ll \sqrt{\nu t}$ (b). See main text for details.}
\label{fig:schematic}
\end{figure}

\begin{figure}[ht]
\includegraphics[width=0.45\textwidth,clip=]{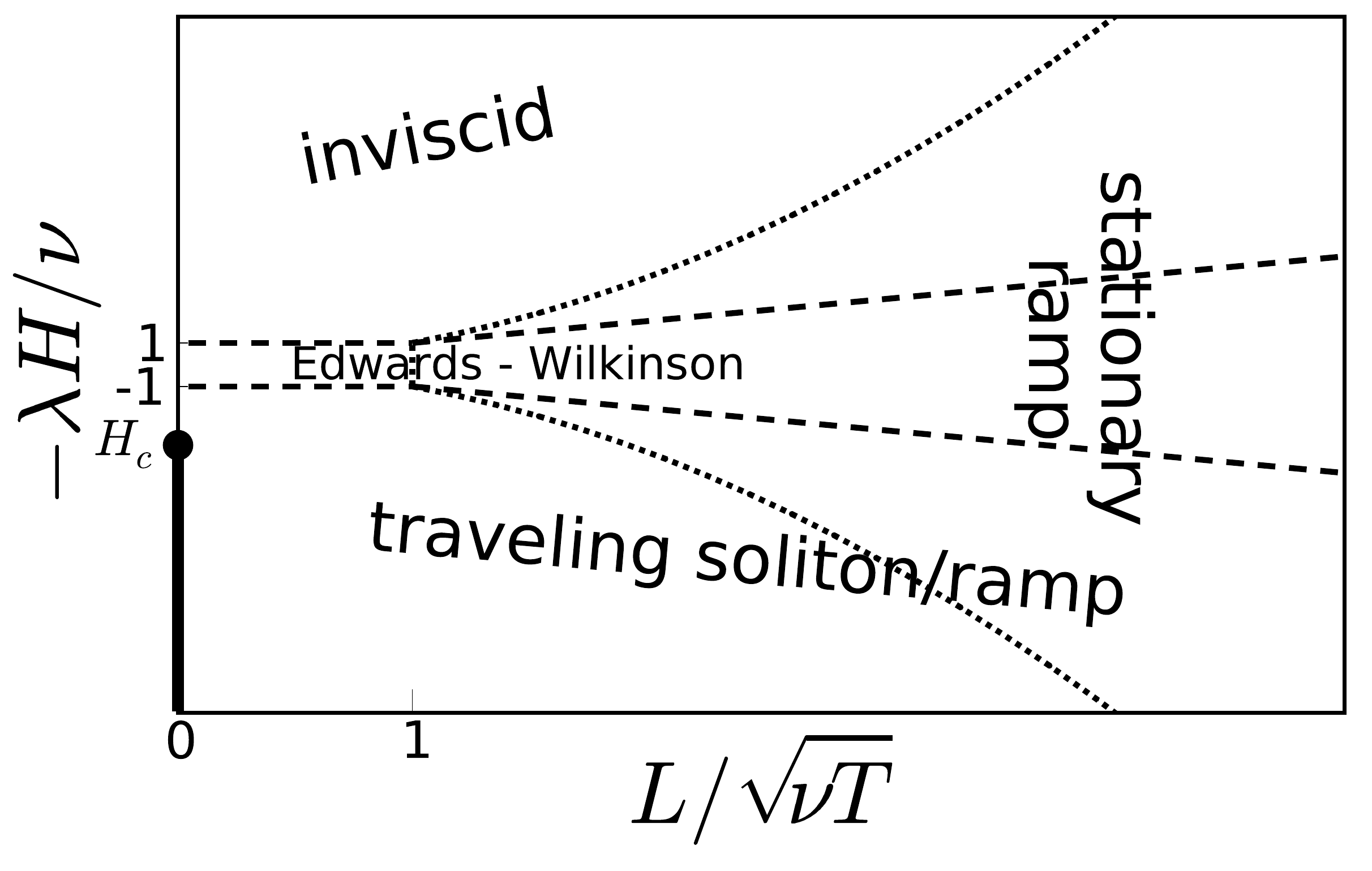}
\caption{Phase diagram in the $\left(L/\sqrt{t},H\right)$ plane. In the stationary ramp, Edwards-Wilkinson (EW), traveling soliton/ramp and inviscid regimes, the height distribution $\mathcal{P}(H,L,t)$ is given by Eqs.~(\ref{eq:stationary_ramp_P}), (\ref{gauss1}), (\ref{s_traveling_wave}), and~(\ref{eq:action_HD_tail}), respectively. The approximate boundaries of the EW and stationary-ramp regimes are denoted by the dashed and dotted lines, respectively.
There is a second-order dynamical phase transition at the point $\left(0,H_{c}\right)=(0,-3.70632489 \dots)$ \cite{Janas2016,LeDoussal2017} as it is crossed in the vertical direction, and a first-order transition when the solid line is crossed. The phase diagram is symmetric with respect to a change of the sign of $L$, so only the regime $L \ge 0$ is shown.}
\label{fig:phase_diagram}
\end{figure}

In this section, we move away from the second-order phase transition and solve the OFM problem~(\ref{eqh})-(\ref{eq:BC_0_and_H}) perturbatively in several regimes: the large $\left|\ell\right|$ limit, the Edwards-Wilkinson regime (which describes \emph{typical} fluctuations of the KPZ interface at short times), and the tails $\lambda H \gg 1$ and  $-\lambda H \gg 1$ for fixed $\ell$. The main results of this section are summarized in Eqs.~(\ref{eq:stationary_ramp_P})-(\ref{eq:action_HD_tail}) above and plotted schematically in Fig.~\ref{fig:schematic}.
A phase diagram of the system in the $\left(L/\sqrt{T},H\right)$ plane is shown in Fig.~\ref{fig:phase_diagram}.

\subsection{Stationary ramp at large $\left|\ell\right|$}
\label{ramp}

The solution is the simplest in the limit where $\left|\ell\right|$ is larger than any dynamical length scale in the problem. Then, in the leading order, the dynamics can be neglected. The optimal profile $h$ is stationary and can be found by minimizing the initial ``cost" $s_{\text{in}}$ over profiles $h(x)$ obeying the constraints $h\left(0\right)=0$ and $h\left(\ell\right)=H$.
This results in a ramp-like profile, which for $\ell>0$ takes the form:
\begin{equation}
\label{eq:stationary_ramp}
h\left(x,t\right)\simeq\begin{cases}
0, & x<0,\\
Hx/\ell, & 0<x<\ell,\\
H, & x>\ell.
\end{cases}
\end{equation}
For $\ell<0$,  all the inequality signs in~(\ref{eq:stationary_ramp}) should be reversed. The action is given, in the leading order, by
$s \simeq s_\text{in} \simeq H^{2}  / \left|\ell\right|$, while $s_\text{dyn}$ is negligible.
The corresponding height-difference distribution~(\ref{eq:stationary_ramp_P}) is Gaussian and
independent of time. This solution is valid for $\left|\ell\right|\gg\max\left\{ 1,\sqrt{\left|H\right|}\right\}$ \citep{footnote:stationary_ramp_validity}.
Eq.~(\ref{eq:stationary_ramp_P}) implies that, for sufficiently large $\left|L\right|$, the probability of observing an unusually large $\left|H\right|$ grows as $\left|L\right|$ is increased. As we observed, this is the case not only in the stationary-ramp regime, but for any $H$ and $\ell$ (at short times that we are dealing with here), see below.

\subsection{Edwards-Wilkinson regime}

\begin{figure}[ht]
\includegraphics[width=0.45\textwidth,clip=]{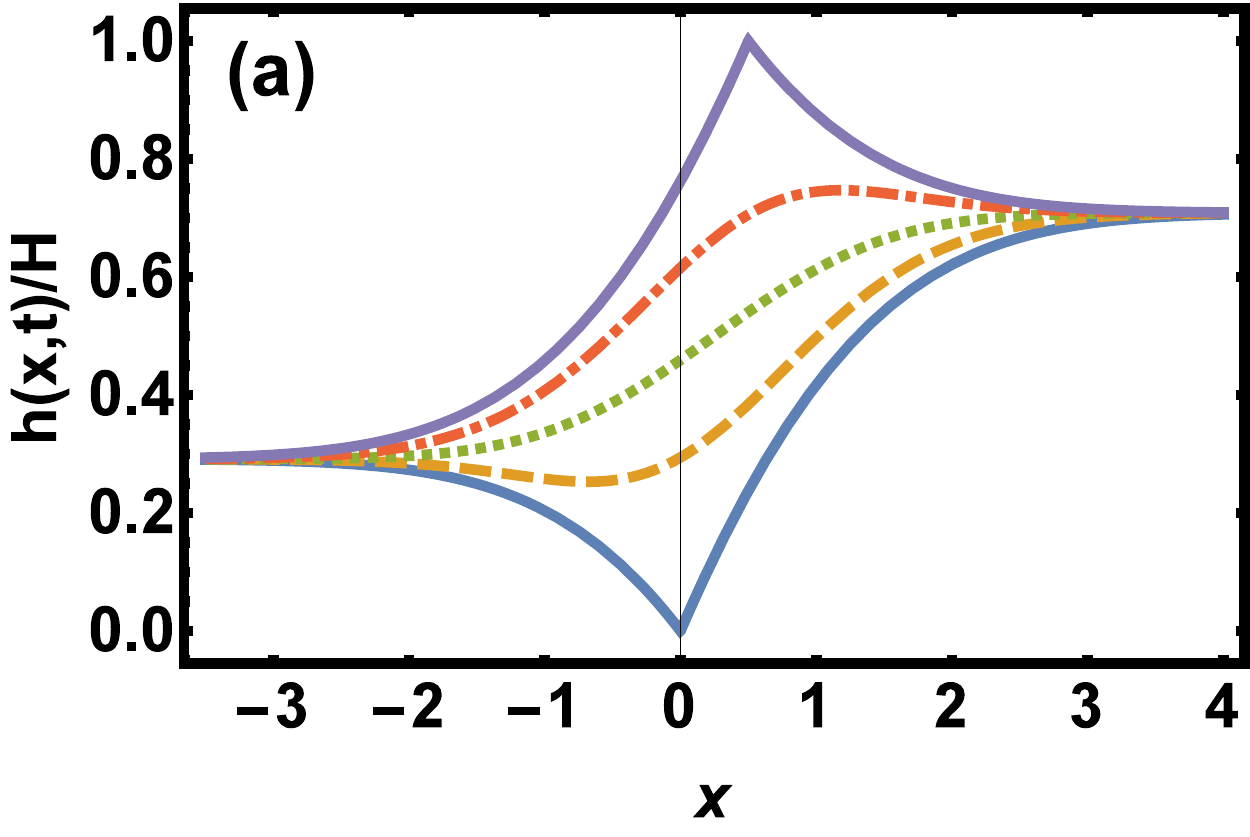}
\includegraphics[width=0.45\textwidth,clip=]{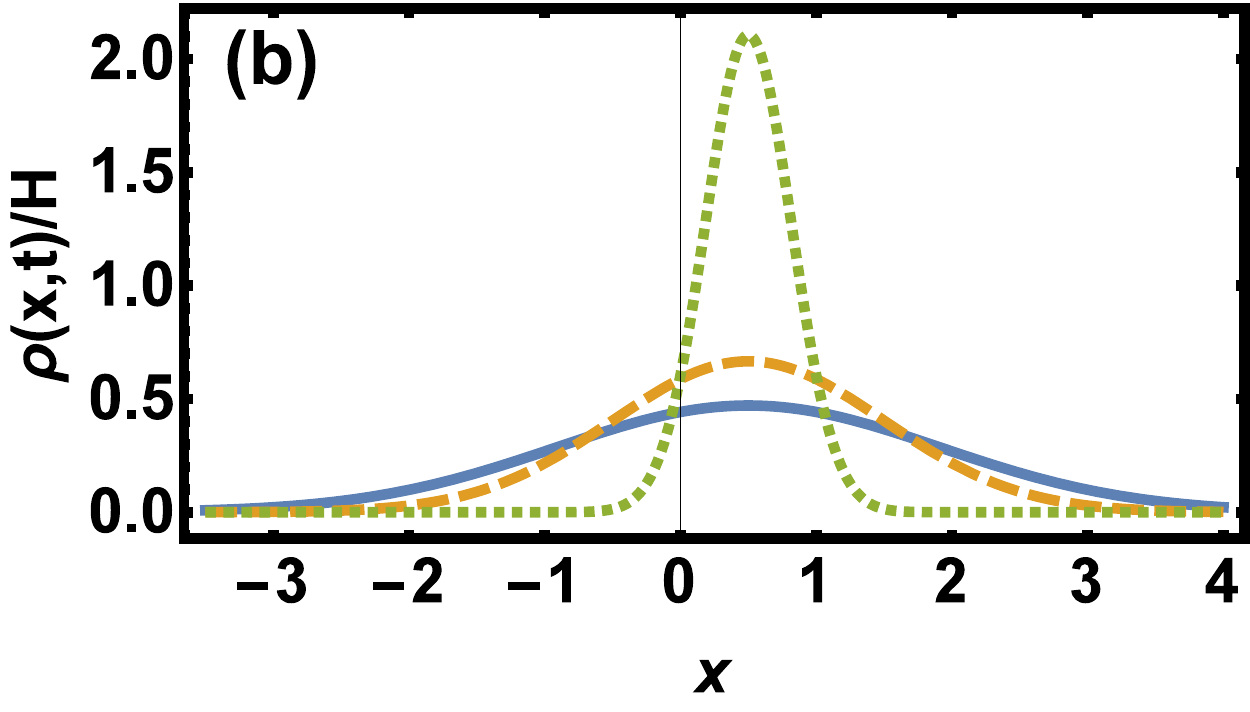}
\caption{The optimal history of the interface (a) and the optimal realization of noise (b) in the EW regime, for $\ell=0.5$, at $t=0,0.25,0.5,0.75$ and $1$ from bottom to top (a) and $t=0,0.5$ and $0.95$ (b).}
\label{fig:EWh}
\end{figure}

For sufficiently small $H$ the OFM problem can be solved via a regular
perturbation expansion in powers of $H$ or $\Lambda_{1}$ \citep{KrMe,MKV,Janas2016}.
One writes $ h(x,t) = \Lambda_{1} h_1(x,t)+\Lambda_{1}^2 h_2(x,t) +\dots $ and similarly for $\rho$.
In the leading order in $\Lambda_{1}$ Eqs.~(\ref{eqh}) and~(\ref{eqrho}) become
\begin{eqnarray}
\label{eq:OFM_EW_h}
\partial_{t}h_{1}     &=& \; \partial_{x}^{2}h_{1}+\rho_{1}, \\
\label{eq:OFM_EW_rho}
\partial_{t}\rho_{1} &=& -\partial_{x}^{2}\rho_{1}.
 \end{eqnarray}
These linear equations correspond to the OFM theory for the Edwards-Wilkinson (EW) equation \citep{EW1982}
\begin{equation}
%\label{eq:EW}
\partial_{t}h=\nu\partial_{x}^{2}h+\sqrt{D}\,\xi(x,t),
\end{equation}
where the KPZ nonlinearity does not play a role. Solving Eq.~(\ref{eq:OFM_EW_rho}) backward in time with initial condition $\rho_{1}\left(x,t=1\right)=\delta\left(x - \ell\right)$, we obtain
\begin{equation}
\label{eq:rho1}
\rho_{1}\left(x,t\right)=G\left(x,\ell,1-t\right)
\end{equation}
where
\begin{equation}
G\left(x,y,t\right)=\frac{1}{\sqrt{4\pi t}}\exp\left[-\frac{\left(x-y\right)^{2}}{4t}\right]
\end{equation}
is the Green's function of the heat equation.
Eqs.~(\ref{eq:OFM_initial_condition}) and~(\ref{eq:rho1}) yield
\begin{equation}
\label{eq:h1double_prime_t_0}
2\partial_{x}^{2}h_{1}\left(x,t=0\right)=\delta\left(x\right)-G\left(x,\ell,1\right).
\end{equation}
Integrating Eq.~(\ref{eq:h1double_prime_t_0}) twice with respect to $x$, and using the conditions $h(x=0,t=0)=0$ and $\partial_{x}h\left(\left|x\right|\to\infty,t=0\right)=0$ we obtain
\begin{eqnarray}
\label{eq:h1_t_0}
&&\antiquad \!\!\!\!\! h_{1}\left(x,t=0\right)=
\frac{e^{-\frac{\ell^{2}}{4}}-e^{-\frac{1}{4}\left(x-\ell\right)^{2}}}{2\sqrt{\pi}}\nonumber\\
&&\;\; +\frac{1}{4}\left[\left|x\right|-(x-\ell)\,\text{erf}\left(\frac{x-\ell}{2}\right)+\ell\,\text{erf}\left(\frac{\ell}{2}\right)\right]
\end{eqnarray}
where $\text{erf}\,z=\left(2/\sqrt{\pi}\right)\int_{0}^{z}e^{-\zeta^{2}}d\zeta$.
We now solve Eq.~(\ref{eq:OFM_EW_h}) and find
\begin{eqnarray}
\label{eq:EW_h1sol}
h_{1}\!\left(x,t\right)&=&\frac{x-\ell}{4}\text{erf}\!\left(\!\frac{\ell-x}{2\sqrt{1-t}}\right)\!+\!\frac{\ell}{4}\text{erf}\left(\!\frac{\ell}{2}\right)\!+\!\frac{x}{4}\text{erf}\left(\!\frac{x}{2\sqrt{t}}\right)\nonumber\\
&+&\frac{1}{2\sqrt{\pi}} \! \left[e^{-\frac{\ell^{2}}{4}}-\sqrt{1-t}\,e^{-\frac{\left(x-\ell\right)^{2}}{4\left(1-t\right)}} \! + \! \sqrt{t}\,e^{-\frac{x^{2}}{4t}}\right] \! ,
\end{eqnarray}
see Fig. \ref{fig:EWh}, yielding
\begin{equation}
\label{eq:H_of_Lambda1_EW}
H=\Lambda_{1}h_{1}\left(\ell,1\right)=\Lambda_{1}\left[\frac{\ell}{2}\text{erf}\left(\frac{\ell}{2}\right)+\frac{e^{-\frac{\ell^{2}}{4}}}{\sqrt{\pi}}\right].
\end{equation}

\begin{figure}[ht]
\includegraphics[width=0.4\textwidth,clip=]{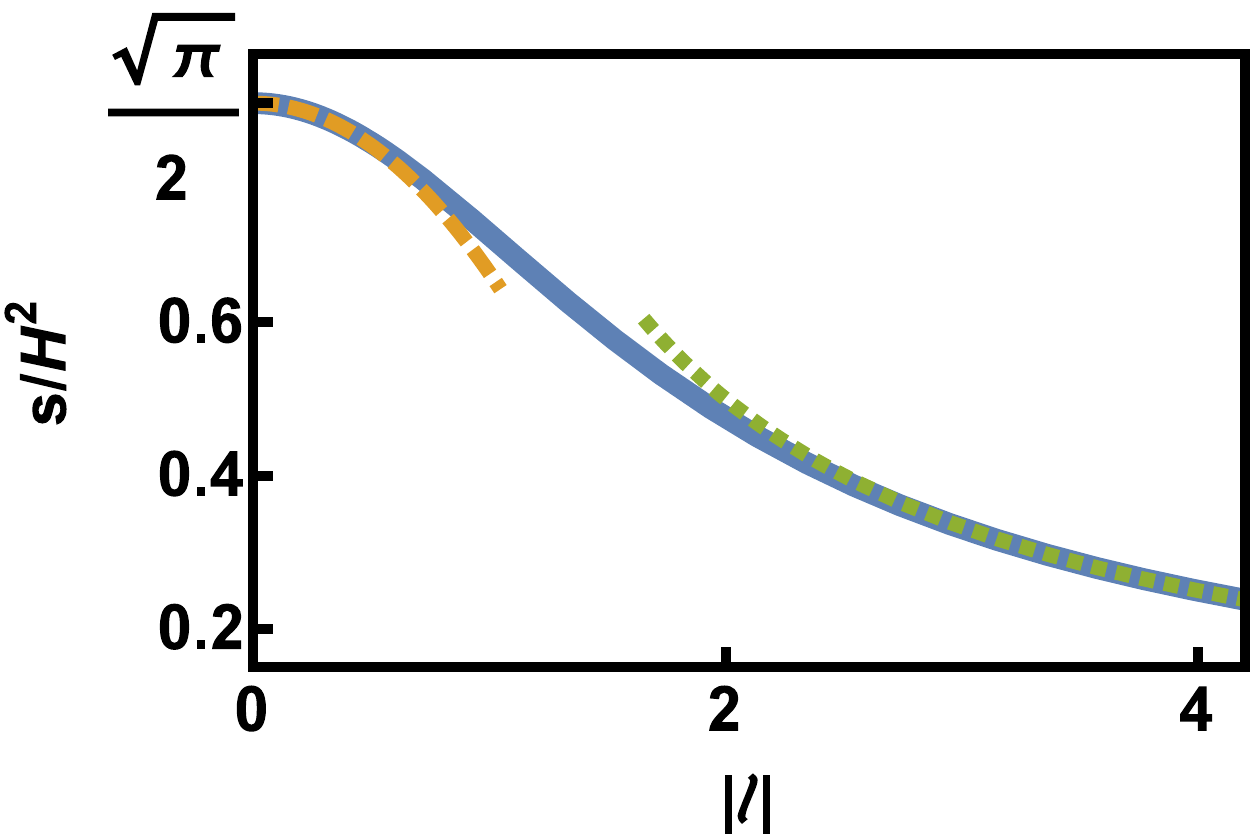}
\caption{The action $s$ vs. $\left|\ell\right|$ in the EW regime as described by the scaling function $g\left(\ell\right)$, see Eqs.~(\ref{gauss1}) and~(\ref{eq:scaling_function_EW}), together with its small- and large-$\left|\ell\right|$ asymptotes (dashed and dotted, respectively).}
\label{fig:EW_s_vs_ell}
\end{figure}

We now evaluate the action:
\begin{eqnarray}
\label{eq:action_EW}
s&=&s_\text{in} + s_\text{dyn}\nonumber\\
&=&\Lambda_{1}^{2}\left\{ \int_{-\infty}^{\infty} \!\!\! dx\left[\partial_{x}h_{1}\left(x,0\right)\right]^{2}+\frac{1}{2}\int_{-\infty}^{\infty} \!\!\! dx \! \int_{0}^{1} \!\! dt\,\rho_{1}^{2}\left(x,t\right)\right\}  \nonumber\\
&=&\Lambda_{1}^{2}\left[\frac{\ell}{4}\text{erf}\left(\frac{\ell}{2}\right)+\frac{e^{-\frac{\ell^{2}}{4}}}{2\sqrt{\pi}}\right].
\end{eqnarray}
Plugging Eq.~(\ref{eq:H_of_Lambda1_EW}) into~(\ref{eq:action_EW}), we obtain Eq.~(\ref{gauss1}) with
\begin{equation}
\label{eq:scaling_function_EW}
g\left(\ell\right)=\left[\ell\,\text{erf}\left(\frac{\ell}{2}\right)+\frac{2e^{-\frac{\ell^{2}}{4}}}{\sqrt{\pi}}\right]^{-1},
\end{equation}
see Fig. \ref{fig:EW_s_vs_ell}.
In the limit $\left|\ell\right|\ll1$, we obtain
$$
s\simeq\frac{\sqrt{\pi}}{2}\left(1-\frac{\ell^{2}}{4}\right) H^2,
$$
which, in the particular case $\ell=0$, reproduces the well-known result \citep{Krug1992,Janas2016}.
Taking the opposite limit $\left|\ell\right| \gg 1$ in Eq.~(\ref{eq:EW_h1sol}),
we find that the optimal profile approaches the stationary ramp~(\ref{eq:stationary_ramp}).
Correspondingly, $s \simeq H^2 / \left|\ell\right|$ in this limit, as we already know from Sec.~\ref{ramp}.
Note that $g$ is a monotonically decreasing function of $\left|\ell\right|$. It follows that the variance of the distribution $\mathcal{P}\left(H,L,T\right)$ increases with $|L|$.

As can be seen from Fig.~\ref{fig:EWh} (a), the optimal profile $h(x,t)$ satisfies the symmetry~(\ref{eq:nontrivial_symmetry}). In addition, it exhibits corner singularities at $x=0$, $t=0$ and at $x=\ell$, $t=1$.

The EW regime requires $\left|\Lambda_{1}\right|\ll1$, or equivalently $\left|H\right|\ll\max\left\{ 1,\left|\ell\right|\right\}$. In the next order of the perturbation expansion in $\Lambda_1$, one can calculate the third cumulant of the height-difference distribution, which already depends on $\lambda$. For flat initial condition such a calculation was performed in Ref. \citep{MKV}.

\subsection{Large positive $\lambda H$}
\label{negtail}

\begin{figure}[ht]
\includegraphics[width=0.4\textwidth,clip=]{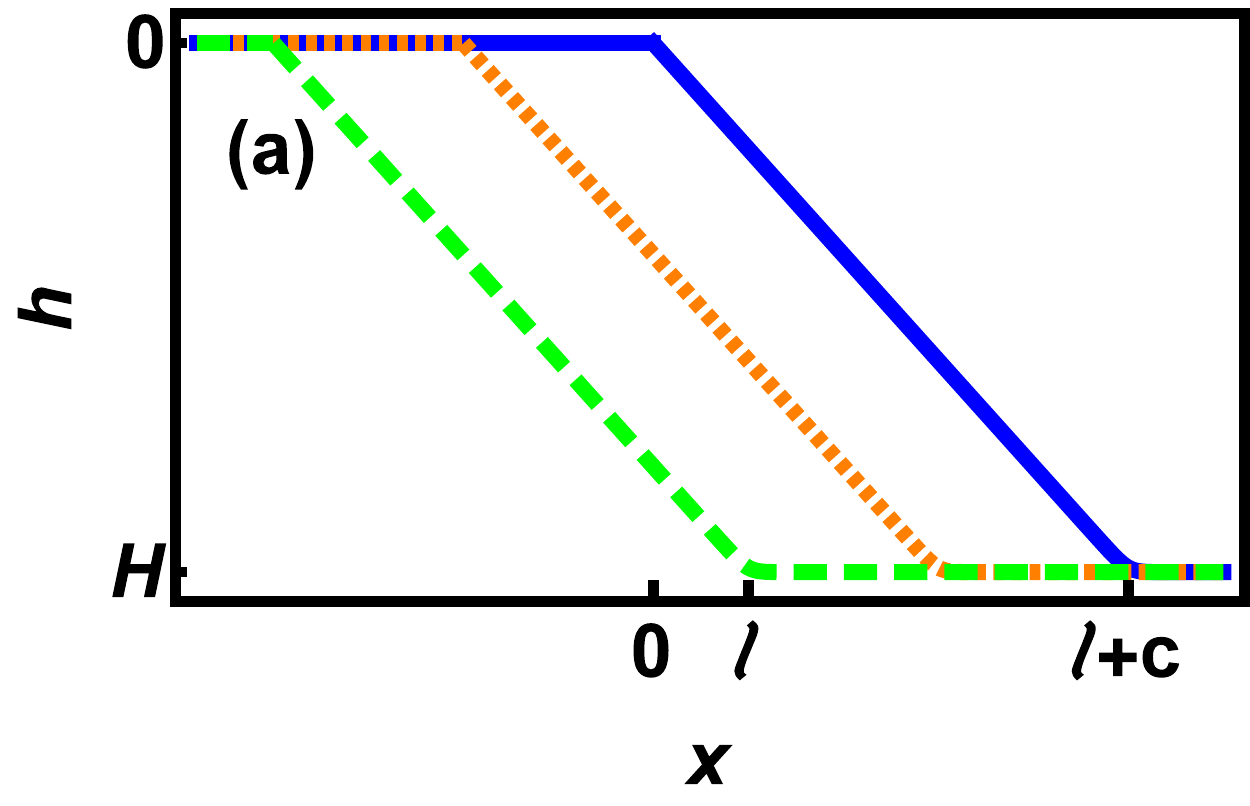}
\includegraphics[width=0.4\textwidth,clip=]{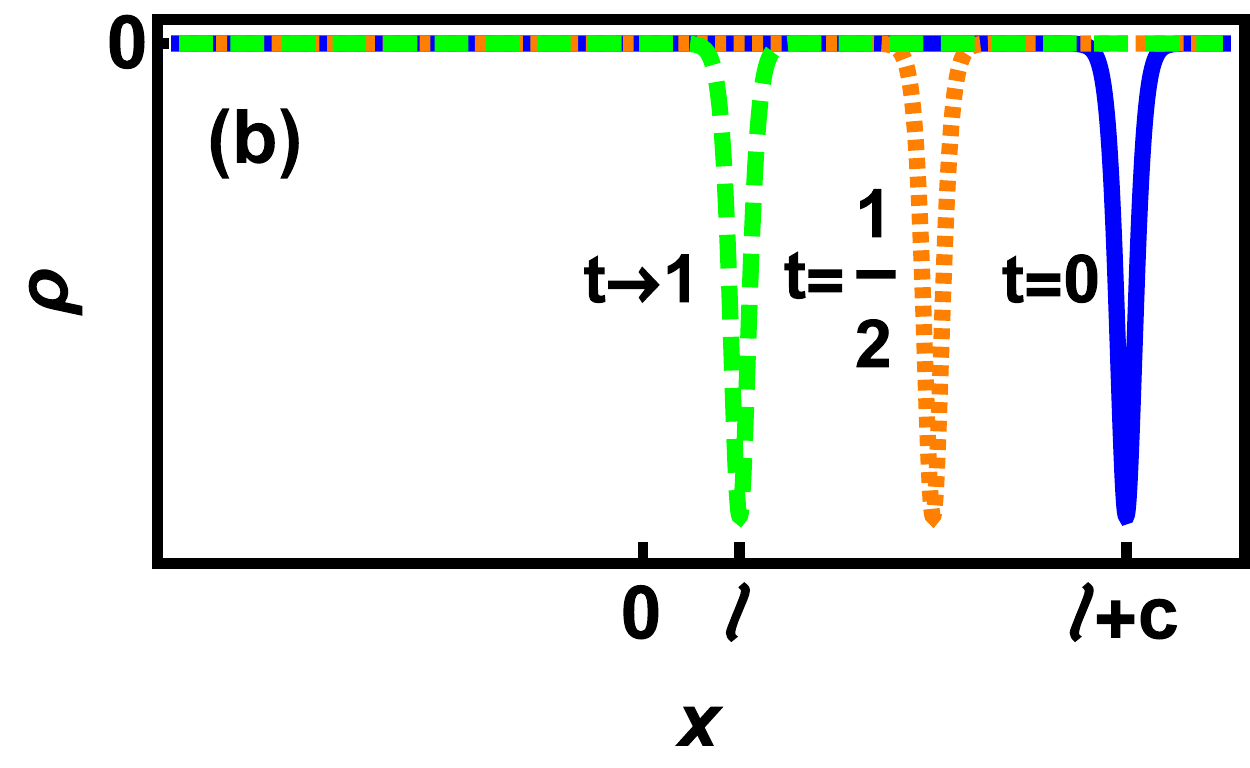}
\caption{The optimal history (a) and optimal realization of the noise (b) in the $\lambda H \gg 1$ tail for $\ell > 0$.  Here $H=-2c\left(\ell+c\right)$.}
\label{fig:negative_tail_h}
\end{figure}

At very large negative $H$, or $\Lambda_{1}$, the optimal solution is provided by one of the two traveling solutions which involve a soliton of $\rho$ and a ``ramp" of $h$. The left-moving solution is given by
\begin{eqnarray}
\label{eq:rho_traveling_wave_ansatz}
\antiquad \rho_{\text{left}}\left(x,t\right)&=&-c^{2}\text{sech}^{2}\left[\frac{c}{2}\left(ct+x-\ell-c\right)\right],\\
\label{eq:h_traveling_wave_ansatz}
\antiquad h_{\text{left}}\left(x,t\right)&\simeq&2\ln\left[1+e^{c\left(ct+x-\ell-c\right)}\right]-2c\left(ct+x\right)
\end{eqnarray}
for $x>-ct$, and $\rho_{\text{left}}\left(x,t\right)\simeq h_{\text{left}}\left(x,t\right)\simeq0$ for $x<-ct$, see Fig.~\ref{fig:negative_tail_h}. These solutions are simple extensions of those obtained in Ref.~\citep{Janas2016} for $\ell =0$, see also Refs.~\citep{Mikhailov1991,Fogedby1998,Fogedby1999,Nakao2003}. Each of these solutions can be also described as a traveling ``shock-antishock" pair of the field
$V(x,t)=\partial_x h$ \citep{Fogedby1998,Fogedby1999,BD2005,BD2006}.

The left-moving solution is optimal (that is, it minimizes the total action $s$) for $\ell>0$.
The optimal solution for $\ell<0$ is a right-moving soliton and ramp $\rho_{\text{right}}\left(x,t\right)$ and $h_{\text{right}}\left(x,t\right)$ respectively, given by the mirror image of Eqs.~(\ref{eq:rho_traveling_wave_ansatz}) and~(\ref{eq:h_traveling_wave_ansatz}) with respect to $x=0$.
The left- and right-moving solutions correspond to the two \emph{local} minima of the Landau free energy $F$ as a function of $\Delta$ in the $H \to -\infty$ limit, see section \ref{sec:DPT} and Fig.~\ref{fig:F_L} (c).

\begin{figure}[ht]
\includegraphics[width=0.4\textwidth,clip=]{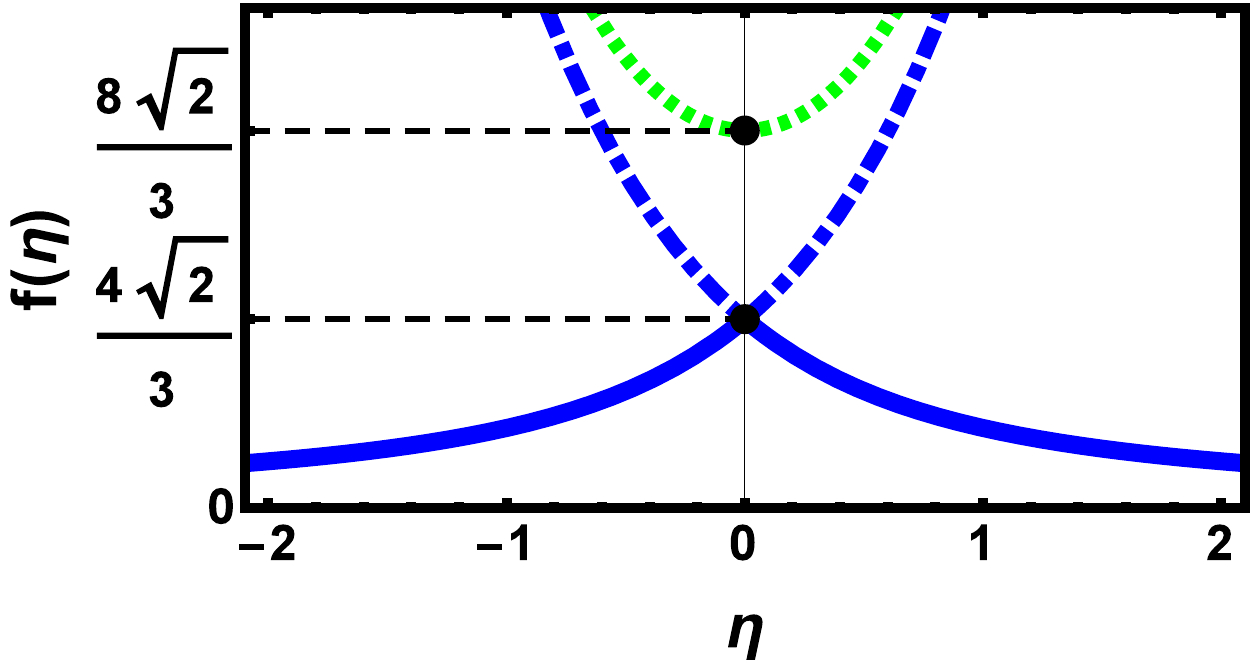}
\caption{Solid line: The scaling function $f(\eta)$ which describes the $\lambda H \gg 1$ tail of the height-difference distribution, see Eqs.~(\ref{s_traveling_wave}) and~(\ref{eq:scaling_function_negative_tail}).
The transition from a smooth curve, similar to that shown in Fig.~\ref{fig:EW_s_vs_ell}, to a curve which exhibits a corner singularity at $\eta=0$ (the solid line in this figure) occurs at $H = H_c$ and has the character of a swallowtail bifurcation \citep{SwallowTail}.
Also shown are the action of the non-optimal traveling ramp solution (dot-dashed) and the action of the non-optimal solution which describes two merging $\rho$-solitons (dotted).}
\label{fig:f_eta}
\end{figure}

The ramp velocity satisfies $c=-\Lambda_{1}/4$  and $s_{\text{dyn}}=4c^{3}/3$ \citep{Janas2016}. Further, $s_{\text{in}}=4c^{2}\left(\ell+c\right)$, while
$H=-2c\left(\ell+c\right)$, so $c$ can be expressed through $H$ and $\ell$:
\begin{equation}
\label{eq:c_ramp}
c=\frac{-\ell+\sqrt{\ell^{2}-2H}}{2}.
\end{equation}
Altogether we find that the action of the left-moving ramp solution is $s\left(H,\ell\right)=\left|H\right|^{3/2}f_{\text{left}}\left(\ell/\sqrt{\left|H\right|}\right)$ with
\begin{equation}
\label{fleft}
f_{\text{left}}\left(\eta\right)=\frac{1}{3}\left(\eta-\sqrt{\eta^{2}+2}\right)^{2}\left(\eta+2\sqrt{\eta^{2}+2}\right),
\end{equation}
and similarly for the right-moving solution with $f_{\text{right}}\left(\eta\right)=f_{\text{left}}\left(-\eta\right)$.
The left- (right-) moving solution is optimal for $\ell > 0$ ($\ell < 0$), resulting in
Eq.~(\ref{s_traveling_wave})
with $f\left(\eta\right)=\min\left\{ f_{\text{left}}\left(\eta\right),f_{\text{right}}\left(\eta\right)\right\}$ given by
 \begin{equation}
 \label{eq:scaling_function_negative_tail}
f\left(\eta\right)=\frac{1}{3}\left(\left|\eta\right|-\sqrt{\eta^{2}+2}\right)^{2}\left(\left|\eta\right|+2\sqrt{\eta^{2}+2}\right),
 \end{equation}
see Fig.~\ref{fig:f_eta}.
The asymptotics of $f\left(\eta\right)$ are
 \begin{equation}
f\left(\eta\right) = \begin{cases}
\frac{4\sqrt{2}}{3}-2\left|\eta\right| + \dots, & \left|\eta\right|\ll1,\\
\frac{1}{\left|\eta\right|}-\frac{1}{3\left|\eta\right|^{3}} + \dots, & \left|\eta\right|\gg1.
\end{cases}
 \end{equation}
Since $f$ is monotonically decreasing with $\left|\eta\right|$, the tail $-H\gg 1$ is enhanced as $\left|L\right|$ is increased.
For $\ell=0$ we obtain $s=4\sqrt{2}\left|H\right|^{3/2}\!\!/3$ in agreement with Refs.
\citep{Janas2016, LeDoussal2017}, and coinciding with the $\lambda H\gg 1$ tail of the Baik-Rains distribution \citep{BR}. At small but nonzero $\ell$  we obtain
\begin{equation}
\label{s_traveling_wave_tail}
s\simeq\left|H\right|^{3/2}\left(\frac{4\sqrt{2}}{3}-2\frac{\left|\ell\right|}{\sqrt{\left|H\right|}}\right).
\end{equation}
At subcritical $H$, the action as a function of $\ell$ is described by a smooth curve which is qualitatively similar to the one shown in Fig. \ref{fig:EW_s_vs_ell}. In contrast, at supercritical $H$, one observes a corner singularity (a jump of $\partial s/\partial\ell$) at $\ell=0$ as in Fig. \ref{fig:f_eta}. This first-order dynamical phase transition is predicted by our Landau theory in Sec. \ref{sec:landau_theory_nonzero_ell}. It has the character of a swallowtail bifurcation as a function
of the parameters $H$ and $\ell$ \citep{SwallowTail}.

In the opposite limit $\left|\ell\right|/\sqrt{\left|H\right|}\gg1$ the velocity of the ramp $c \simeq -H / (2\ell)$ is relatively small: $\left|c\right|\ll\left|\ell\right|$.  In the leading order, the ramp~(\ref{eq:stationary_ramp}) does not move at all, leading to the action $s\simeq s_{\text{in}}\simeq H^{2}/\left|\ell\right|$. In the subleading order we obtain
 \begin{equation}
 \label{largeell1}
s\simeq\frac{H^{2}}{\left|\ell\right|}\left(1-\frac{\left|H\right|}{3\ell^{2}}\right).
 \end{equation}

As we mentioned earlier, the optimal profile obeys the combined symmetry~(\ref{eq:nontrivial_symmetry}). This symmetry is evident in Fig.~\ref{fig:negative_tail_h} (a). For $\ell=0$ this is the case even though the optimal profile is not mirror-symmetric in space.%, that is, $h(x,t) \ne h(-x,t)$.

\begin{figure}[ht]
\includegraphics[width=0.4\textwidth,clip=]{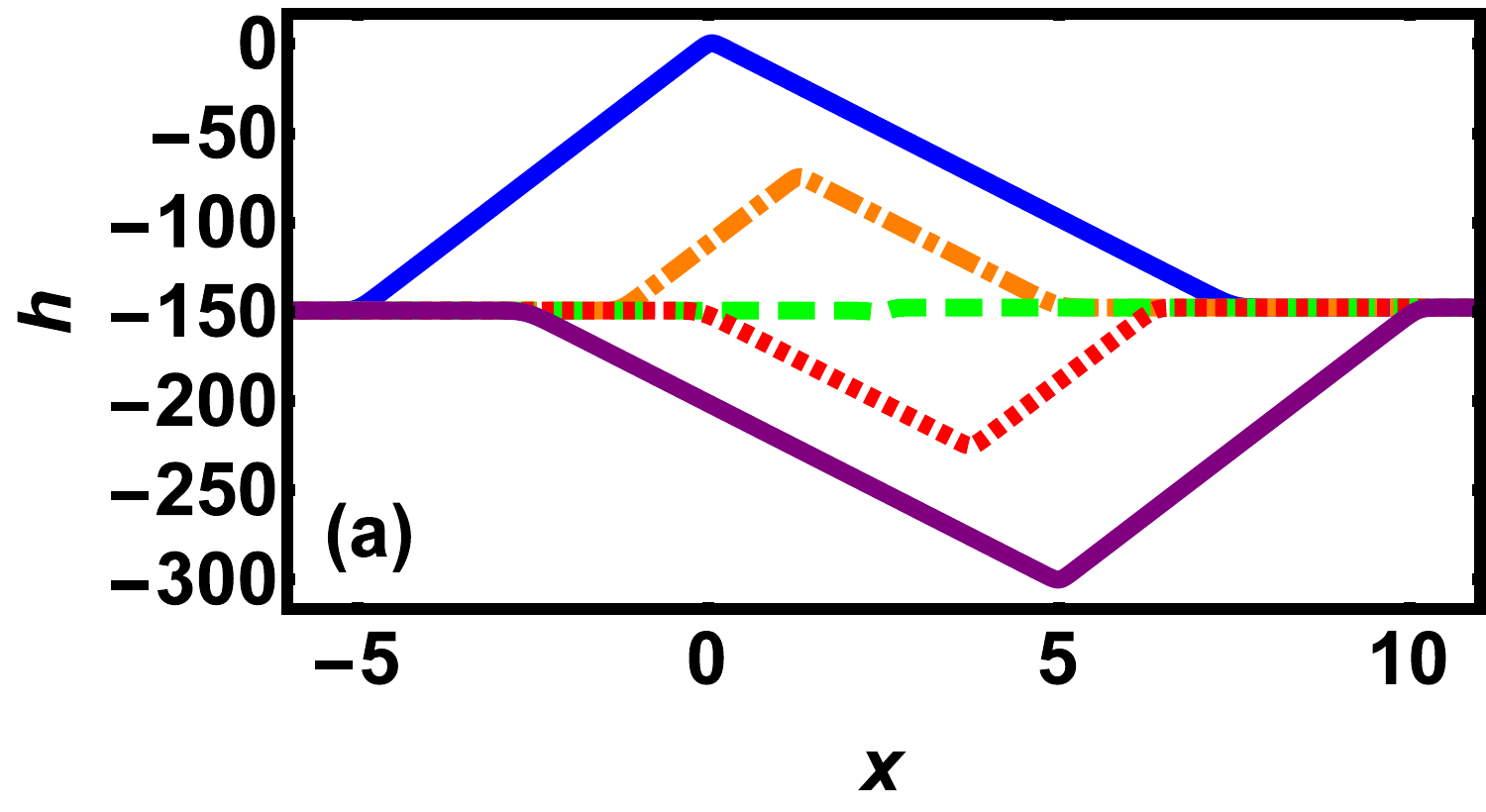}
\includegraphics[width=0.4\textwidth,clip=]{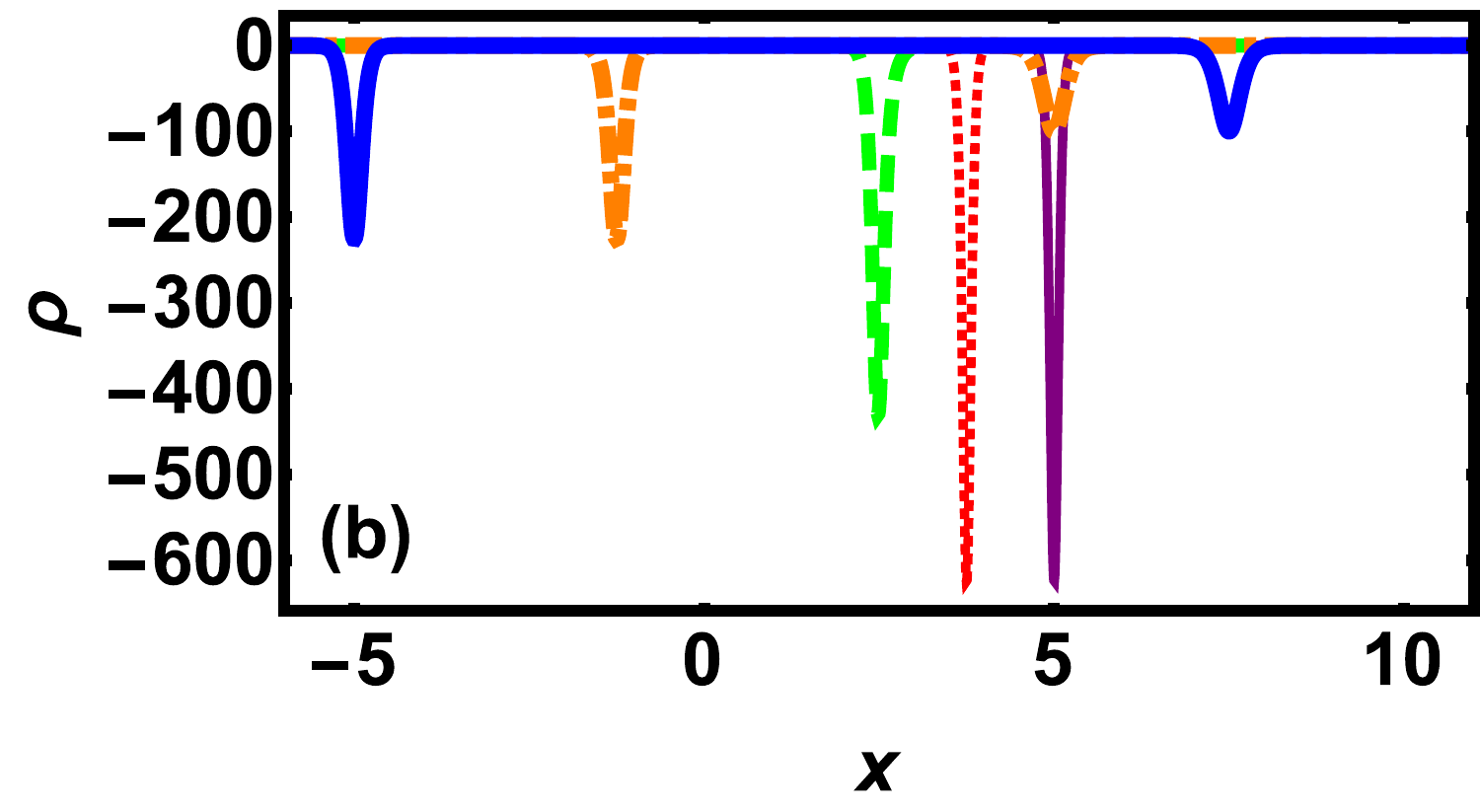}
\caption{Example of exact solutions~(\ref{eq:merging_solitons_h}) and~(\ref{eq:merging_solitons_rho}) which describe merger of two different counter-propagating $\rho$-solitons and subsequent motion of a single soliton.  Here $N=3$, $c_1 = X_1 = 0$, $c_2=-10$, $c_3 = 15$, $X_{2}=c_{3}/2$ and $X_{3}=c_{2}/2$.
Shown are (a) $h$ vs. $x$ at times $t=0, 1/4, 1/2, 3/4$ and $1$ from top to bottom and (b) $\rho$ vs. $x$ at times $t=0$ (solid, thick), $1/4$ (dot-dashed), $1/2$ (dashed), $3/4$ (dotted) and $1$ (solid, thin).
In the limit $\lambda H \gg 1$, this type of solution describes the \emph{non-optimal} third solution $\left(h_{\text{m}},\rho_{\text{m}}\right)$ to the OFM problem, with $\ell = c_2 + c_3$ and $H = 2c_2c_3$. The action of this non-optimal solution is given by Eq.~(\ref{thirdaction}) and shown by the dotted line in  Fig.~\ref{fig:f_eta}.}
\label{fig:third_sol}
\end{figure}

In addition to the left- and right-moving ramp solutions, there is a third (non-optimal) solution $\rho_{\text{m}}(x,t)$ and $h_{\text{m}}(x,t)$. It describes a collision and
merger, at $t=1/2$ and $x=\ell/2$, of \emph{two} different oppositely moving $\rho$-solitons,  see Fig.~\ref{fig:third_sol}.
The merger is mediated by the ordinary ($\rho=0$) shock of $V = \partial_x h $ which starts at $t=0$ at $x=0$ and arrives, at $1/2$, at the same point $x=\ell/2$ as the two solitons. Upon merger a single traveling soliton is formed which arrives at the point $x=\ell$ at  $t=1$. This solution corresponds to the local \emph{maximum} of $F(\Delta)$ at $H \to -\infty$, see Fig. \ref{fig:F_L} (c).
Remarkably, this solution belongs to the family of exact multi-soliton solutions of Eqs.~(\ref{eqh}) and~(\ref{eqrho}), discovered in Ref. \citep{Janas2016}:
\begin{eqnarray}
\label{eq:merging_solitons_h}
&h\left(x,t\right) = 2\ln\left[\frac{C\sum_{i=1}^{N}e^{c_{i}\left(c_{i}t-x+X_{i}\right)}}{\sum_{i,j=1}^{N}\left(c_{i}-c_{j}\right)^{2}e^{c_{i}\left(c_{i}t-x+X_{i}\right)+c_{j}\left(c_{j}t-x+X_{j}\right)}}\right],\nonumber\\\\
\label{eq:merging_solitons_rho}
&\rho\left(x,t\right) = -\frac{2\sum_{i,j=1}^{N}\left(c_{i}-c_{j}\right)^{2}e^{c_{i}\left(c_{i}t-x+X_{i}\right)+c_{j}\left(c_{j}t-x+X_{j}\right)}}{\left[\sum_{i=1}^{N}e^{c_{i}\left(c_{i}t-x+X_{i}\right)}\right]^{2}}.\nonumber\\
\end{eqnarray}
The particular case $N=3$, $c_1 = X_1 = 0$,
\begin{equation*}
c_{2}=\frac{\ell-\sqrt{\ell^{2}-2H}}{2},\quad c_{3}=\frac{\ell+\sqrt{\ell^{2}-2H}}{2},
\end{equation*}
$X_{2}=c_{3}/2$ and $X_{3}=c_{2}/2$ approximately satisfies all of the boundary conditions in the $\lambda H\to \infty$ limit. It also obeys the combined symmetry~(\ref{eq:nontrivial_symmetry}), see Fig.~\ref{fig:third_sol}. The arbitrary constant $C$ can be chosen so that $h(0,0)=0$.

We will skip a more detailed description of this beautiful but non-optimal (and, therefore, non-physical) solution and confine ourselves to presenting its action:
\begin{equation}\label{thirdaction}
s=\left|H\right|^{3/2}f_{\text{m}}\left(\ell/\sqrt{\left|H\right|}\right),\quad
f_{\text{m}}\left(\eta\right)=\frac{4}{3}\left(\eta^{2}+2\right)^{3/2} ,
\end{equation}
see Fig.~\ref{fig:f_eta}. Interestingly, it is equal to the sum of the actions of the other two solutions: $f_{\text{m}}\left(\eta\right) = f_{\text{left}}\left(\eta\right)+f_{\text{right}}\left(\eta\right)$, where the functions
$f_{\text{left}} \left(\eta\right)$ and  $f_{\text{right}} \left(\eta\right)$ were defined in Eq.~(\ref{fleft}) and in the subsequent paragraph.   Moreover, the velocities $c_2$ and $c_3$ of the two merging solitons are equal to the velocities of the one-soliton solutions $\rho_{\text{left}}$ and $\rho_{\text{right}}$, respectively. Finally, $h_{\text{m}}$ has the property $h_{\text{m}}\left(\left|x\right|\to\infty,t\right)\simeq H/2$.

In the particular case $\ell = 0$, the solution $\left(h_{\text{m}},\rho_{\text{m}}\right)$ is symmetric with respect to $x \leftrightarrow -x$, and its action is $s(H)=8\sqrt{2}\,\left|H\right|^{3/2}\!/3$. As observed in Ref.~\cite{Janas2016}, this action coincides with the corresponding tail of the Tracy-Widom distribution \citep{TracyWidom1994} which is non-optimal for stationary interface. As observed in Ref.~\citep{LeDoussal2017}, this tail is described by the $\lambda H\to \infty$ asymptote of a \emph{non-physical} branch obtained via analytical continuation of the exact subcritical  large-deviation function at short times.  The correct branch is obtained via a \emph{non-analytic} continuation \citep{LeDoussal2017}.

The results of this subsection are valid for ramp velocities~(\ref{eq:c_ramp}) much larger than unity, or equivalently for $-H \gg\max\left\{ \left|\ell\right|,1\right\}$.

\subsection{Large negative $\lambda H$}
\label{sec:hydrodynamics}

In this regime the optimal path is large-scale in terms of both $h$ and $\rho$, and one can neglect the diffusion terms
in Eqs.~(\ref{eqh}), (\ref{eqrho}) and~(\ref{eq:OFM_initial_condition}). The resulting problem
is mappable into a one-dimensional inviscid hydrodynamics of a compressible ``gas" with density $\rho\left(x,t\right)$ and velocity $V\left(x,t\right) =\partial_x h \left(x,t\right)$ \cite{MKV}: %Indeed, Eqs.~(\ref{eqh}) and (\ref{eqrho}) become
\begin{eqnarray}
 \partial_t \rho +\partial_x (\rho V)&=& 0, \label{rhoeq}\\
  \partial_t V +V \partial_x V &=&\partial_x \rho, \label{Veq}
\end{eqnarray}
This ``gas" has \emph{negative} pressure $p\left(\rho\right)=-\rho^2/2$.
The problem should be solved subject to the boundary conditions
\begin{equation}
\label{eq:boundary_condition_HD}
\rho\left(x,t=0\right)=\Lambda_{1}\delta\left(x\right),\quad\rho\left(x,t=1\right)=
\Lambda_{1}\delta\left(x-\ell\right).
\end{equation}
Since diffusion is neglected, so must be $s_\text{in}$ \citep{Janas2016}, therefore $s = s_{\text{dyn}}$ up to subleading corrections.

The solution to the problem~(\ref{rhoeq})-(\ref{eq:boundary_condition_HD}) for $\ell \neq 0$ can be
expressed, by using a Galilean transformation,  through the solution of the same problem with $\ell=0$.
The latter solution was obtained in Ref.~\citep{Janas2016}, see also Ref.~\citep{KMSparabola}. In the ``pressure"-dominated region $\left|x-\ell t\right|\le\mathfrak{L}\left(t\right)$,
the solution for $\ell \neq 0$ can be written as
\begin{eqnarray}
\label{eq:rho_HD_galilean}
\rho\left(x,t\right)&=&\rho_{0}\left(x-\ell t,t\right),\\
\label{eq:V_HD_galilean}
V\left(x,t\right)&=&V_{0}\left(x-\ell t,t\right)+\ell,
\end{eqnarray}
where
\begin{equation}
\label{eq:V0sol}
V_{0}\left(y,t\right)=-a\left(t\right)y
\end{equation}
and
\begin{equation}
\rho_{0}\left(y,t\right)=r\left(t\right)\left[1-\frac{y^{2}}{\mathfrak{L}^{2}\left(t\right)}\right]
\end{equation}
are the uniform-strain flow solutions for $\ell=0$ \citep{KMSparabola,Janas2016}. The functions $a\left(t\right)$, $\mathfrak{L}\left(t\right)$ and $r\left(t\right)$ were determined in Ref.~\citep{KMSparabola}.
As shown below, the action, which we now calculate in terms of $H$ and $\ell$, is completely determined by pressure-dominated region.

Equations~(\ref{eq:rho_HD_galilean}) and~(\ref{actiongeneral}) imply that the action, expressed via the Lagrange multiplier $\Lambda_1$, does not depend on $\ell$. That is, $s\left(\Lambda_{1},\ell\right)\simeq s_{0}\left(\Lambda_{1}\right)$
where $s_{0}\left(\Lambda_{1}\right)\simeq\left(3\pi\right)^{2/3} \! \Lambda_{1}^{5/3} \! /5$, the action for $\ell=0$, was found in Ref.~\citep{Janas2016}.
For $\ell=0$, the calculation proceeds as follows \citep{KMSparabola}. Neglecting the diffusion term in Eq.~(\ref{eqh}) and using $V_{0}\left(0,t\right)=0$ we obtain
\begin{eqnarray}
\label{eq:H_0_of_Lambda1}
H_{0}\left(\Lambda_{1}\right)&=&h\left(0,1\right)-h\left(0,0\right)=\int_{0}^{1}dt\,\partial_{t}h\left(0,t\right)=\nonumber\\
&=&\int_{0}^{1}\rho_{0}\left(0,t\right)dt=\int_{0}^{1}r\left(t\right)dt.
\end{eqnarray}
Evaluating the integral~(\ref{eq:H_0_of_Lambda1}) one obtains $H_{0}\left(\Lambda_{1}\right)\simeq\left(3\pi\Lambda_{1}\right)^{2/3}\!/2$, leading to $s_{0}\left(H_{0}\right)=4\sqrt{2}\,H_{0}^{5/2}\!/\left(15\pi\right)$ \citep{KMSparabola,Janas2016}.

For $\ell \ne 0$, we again neglect the diffusion term in Eq.~(\ref{eqh}), and then use Eqs.~(\ref{eq:rho_HD_galilean}) and~(\ref{eq:V_HD_galilean}) in order to obtain
\begin{eqnarray}
\label{eq:H_of_Lambda1_HD}
H\!&=&h\left(x=\ell,t=1\right)-h\left(x=0,t=0\right)\nonumber\\
&=&\int_{0}^{1}dt\frac{d}{dt}\left[h\left(x=\ell t,t\right)\right]\nonumber\\
&=&\int_{0}^{1}dt\left[\ell\partial_{x}h\left(x=\ell t,t\right)+\partial_{t}h\left(x=\ell t,t\right)\right]\nonumber\\
& \simeq&\! \int_{0}^{1} \!\!\!\! dt\left[\ell V\left(x=\ell t,t\right)-\frac{1}{2}V^{2}\left(x=\ell t,t\right)+\rho\left(x=\ell t,t\right)\right]\nonumber\\
&=&\int_{0}^{1}dt\left[\ell\left(V_{0}+\ell\right)-\frac{1}{2}\left(V_{0}+\ell\right)^{2}+\rho_{0}\right]_{x=0}\nonumber\\
&=&\frac{\ell^{2}}{2}+H_{0}\left(\Lambda_{1}\right).
\end{eqnarray}
As a result, $s\left(H,\ell\right)\simeq s_{0}\left(H_0 = H-\ell^{2}/2\right)$, which yields Eq.~(\ref{eq:action_HD_tail}).
$s\left(H,\ell\right)$ is a monotonically decreasing function of $\left|\ell\right|$, implying that as $\left|L\right|$ is increased, it becomes more likely to observe an unusually large positive $H$.

For $\left|x-\ell t\right| > \mathfrak{L}\left(t\right)$, $\rho(x,t)$ vanishes, so this region does not contribute to the action. Here
$V(x,t)$ satisfies the Hopf equation
\begin{equation}\label{Hopfeq}
\partial_tV+V\partial_xV=0 .
\end{equation}
The solution of this equation should be continuously matched, at $\left|x-\ell t\right| = \mathfrak{L}\left(t\right)$,  with the pressure-driven solution. It should also obey the boundary conditions $V\left(x\to\pm\infty,t\right)=0$.
In the particular case $\ell=0$, $V(x,t)$ must respect the symmetries $V_{0}\left(x,t\right)=-V_{0}\left(-x,t\right)=-V_{0}\left(x,1-t\right)$, which are directly related to the spatial mirror symmetry of the OFM problem and to the symmetry~(\ref{eq:nontrivial_symmetry}).
These symmetries cannot be spontaneously broken, as otherwise a dynamical phase transition would occur at some value $\lambda H < 0$. Such a transition, however, is impossible, because  the exact short-time large-deviation function for $\ell = 0$ is known to be analytic at all $\lambda H < 0$ \citep{LeDoussal2017}.

The symmetry $V_{0}\left(x,t\right)=-V_{0}\left(x,1-t\right)$ \citep{SmithMeerson2018} and the exact short-time results for $\mathcal{P}\left(H,L=0,t\right)$ \citep{LeDoussal2017} have been uncovered very recently. They were unknown to the authors of Ref. \citep{Janas2016}, and this led to a mistake in their Hopf-flow solution (see
Fig.~8 of their Appendix C). Although this mistake did not affect the action, for completeness we now present the correct Hopf solution for $\ell = 0$.

\begin{figure}[ht]
\includegraphics[width=0.4\textwidth,clip=]{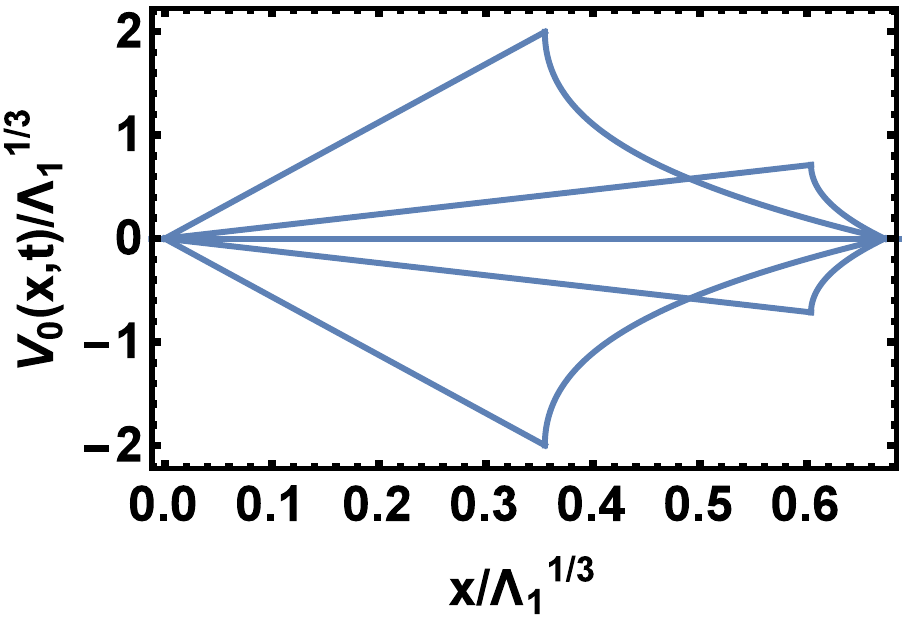}
\caption{$V(x,t)=\partial_{x}h(x,t)$ as a function of $x$ for large negative $\lambda H$ and $\ell=0$ at times $t=0.1$, $0.3$, $0.5$, $0.7$ and $0.9$, from top to bottom. The solution respects the symmetries $V_{0}\left(x,t\right)=-V_{0}\left(-x,t\right)=-V_{0}\left(x,1-t\right)$.}
\label{fig:V_of_x_HD}
\end{figure}

In the Hopf region $\left|x\right|>\mathfrak{L}\left(t\right)$
there are multiple solutions to Eq.~(\ref{Hopfeq}) which can be continuously matched to the pressure-dominated region while satisfying the boundary conditions at $x \to \pm \infty$ via a weak discontinuity or a shock. The ensuing selection problem is a price to pay for the inviscid approximation: as argued above, with account of diffusion the OFM problem has a unique solution in the $H \gg 1$ tail, and it must respect all of the symmetries of the problem.
Imposing the symmetry~(\ref{eq:nontrivial_symmetry}), we now construct the correct solution at $0 \le t \le 1/2$ from the known solution at $1/2 \le t \le 1$ \citep{KMSparabola}.
In the Hopf region $V_{0}\left(x,t\right)$ is given in terms of $\tilde{x}=x/\Lambda_{1}^{1/3}$ and $\tilde{V}_{0}(\tilde{x},t)=V_{0}(x,t)/\Lambda_{1}^{1/3}$ as follows.
In the region $\left|\tilde{x}\right| > 3/\left(4r_{*}\right)$,
where $r_{*}=(3\pi)^{2/3} \! /4$,
$\tilde{V}_0(x,t)$
vanishes. For $\left|\tilde{x}\right| < 3/\left(4r_{*}\right)$, the solution is given
by the algebraic equation
\begin{eqnarray}
&&\antiquad \antiquad \text{sgn}\left(\! t-\frac{1}{2} \right)\tilde{x}-\tilde{V}_{0}\psi\left(t\right)=-\frac{\tilde{V}_{0}}{2}
\nonumber\\
&&- \, \text{sgn}\left(\tilde{V}_{0}\right)\left[\frac{\tilde{V}_{0}}{\pi}
\text{arctan}\left(\frac{\tilde{V}_{0}}{2\sqrt{r_{*}}}\right)+\frac{3}{4r_{*}}\right] ,
\end{eqnarray}
where
\begin{equation}
\psi\left(t\right)=\begin{cases}
1-t, & 0\leq t\leq\frac{1}{2},\\
t, & \frac{1}{2}\leq t\leq 1 .
\end{cases}
\end{equation}
This solution, alongside with its counterpart~(\ref{eq:V0sol}) in the ``pressure"-dominated region, is presented in Fig.~\ref{fig:V_of_x_HD}.
$V\left(x,t\right)$ can be integrated with respect to $x$ to yield $h\left(x,t\right)$. The solution can also be found for nonzero $\ell$. We do not show these cumbersome calculations because they do not contribute to the action in the leading order we are after.
However, we will comment on one interesting feature of the solution. In the inviscid limit, the $h$-profile exhibits \emph{cusp} singularities at $x=\ell$, $t=1$ and at $x=0$, $t=0$. Diffusion partially smoothes these singularities, so that only corner singularities remain.
Using the symmetry~(\ref{newsymmetry}), one can show that $h(x,t=1)$ must exhibit a corner singularity at the single point where the height is measured. This is true for any initial condition \citep{SmithMeerson2018}, as indeed exemplified by all known particular cases \citep{MKV, KMSparabola, Janas2016, MeersonSchmidt2017}.

Finally, we check the conditions for the strong inequality $s_{\text{in}}\ll s_{\text{dyn}}$, assumed in this subsection, by comparing the action~(\ref{eq:action_HD_tail}) with that of the stationary ramp solution ($s\simeq s_{\text{in}}\simeq H^{2}\!/\left|\ell\right|$).
We find that the results of this subsection are valid for $H-\ell^2/2 \gg\max\left\{ 1,\left|\ell\right|^{6/5}\right\}$.

\section{Summary and discussion}

\label{disc}

In this paper, we built on the results of Janas \textit{et al.} \citep{Janas2016} and Krajenbrink and Le Doussal \citep{LeDoussal2017},
who studied the distribution $\mathcal{P}\left(H,t\right)$ of the two-time height difference $H$ of a stationary 1d KPZ interface at short times, using the OFM and the exact representation \citep{IS, Borodinetal}, respectively.
We focused our attention on the second-order dynamical phase transition -- a singularity of the large deviation function -- at $\lambda H=\lambda H_c > 0$, caused by spontaneous breaking of the reflection symmetry by the optimal path leading to a given $H$ \cite{Janas2016}.

We developed an effective Landau theory of the second-order phase transition by defining a proper order parameter~(\ref{eq:Delta_def}) which quantifies the spatial reflection asymmetry of the optimal interface at $t=0$.
Here the large deviation function of the distribution $s=-\epsilon\ln\mathcal{P}$ and $\lambda H$ play the roles of equilibrium free energy and inverse temperature, respectively.

We also generalized the problem by considering the distribution of the two-time height difference between two points at distance $L$ apart.
We found that, near the critical point $H=H_c$, $L / \sqrt{t}$ plays the role of external magnetic field in the traditional Landau theory.
The nonequilibrium analog of the Landau theory, formulated here,  yields critical exponents which provide a detailed characterization of the singularities of $s$ and of the order parameter $\Delta$ at the critical point.
In particular, we found that at supercritical $H$, a change of the sign of $L$ is accompanied by a first-order dynamical phase transition. This transition has the character of a swallowtail bifurcation.

In addition, we evaluated $\mathcal{P}\left(H,L,t\right)$ analytically in several limits away from the second-order phase transition by finding perturbative solutions to the OFM problem, see Fig.~\ref{fig:schematic}. Our asymptotic results for $\mathcal{P}\left(H,L,t\right)$ are given by Eqs.~(\ref{gauss1}) and~(\ref{eq:scaling_function_EW}) for small fluctuations,
by Eqs.~(\ref{s_traveling_wave}) and~(\ref{eq:scaling_function_negative_tail}) for large positive $\lambda H$,
and by Eq.~(\ref{eq:action_HD_tail}) for large negative $\lambda H$.
In the large-$\left|L\right|/\sqrt{t}$ limit, $\mathcal{P}$ is given by Eq.~(\ref{eq:stationary_ramp_P}).
We observed that $s$ is a monotonically decreasing function of $\left|L\right|/\sqrt{t}$, implying that increasing $\left|L\right|$ facilitates large deviations of $H$.
In analogy with other initial conditions \citep{MKV,KMSparabola}, we expect the $\lambda H \gg 1$ tail~(\ref{s_traveling_wave}) %(\ref{s_traveling_wave_tail}) {s_traveling_wave}
to hold at arbitrary times.

The optimal initial condition $h(x,t=0)$ which leads to a given $H$ figures prominently in the solution to the OFM problem~(\ref{eqh})-(\ref{eq:BC_0_and_H}). An interesting question is how this initial condition is created at earlier,  ``pre-historic" times, $t<0$.  We address this question in the Appendix.

Recently Le Doussal \citep{LeDoussal2017combined} used the replica Bethe ansatz to obtain an exact representation for the distribution of the one-point, two-time height difference $h\left(0,t\right)-h\left(0,0\right)$ for the KPZ interface for a combined initial condition which is flat at $x<0$ and stationary at $x>0$.
This problem can be generalized by considering the distribution $\mathcal{P}\left(H,L\right)$ of the two-point, two-time height difference $H=h\left(L,t\right)-h\left(0,0\right)$ with the same combined initial condition.
It so happens that, for very large positive $\lambda H$ and $L \ge 0$, the interface history~(\ref{eq:h_traveling_wave_ansatz}) satisfies the condition $h\left(x<0,t=0\right)=0$.
It is therefore the optimal history for the combined initial condition of Ref.~\citep{LeDoussal2017combined} at $L\ge 0$. As a result, the $\lambda H\gg 1$ tail
of  $\mathcal{P}\left(H,L\right)$ for the combined initial condition is given by Eqs.~(\ref{s_traveling_wave}) and~(\ref{eq:scaling_function_negative_tail}) for $L\ge 0$,
whereas the deterministic part, $x<0$, of the initial condition, does not contribute in the leading order.
For $L=0$ this tail coincides with the corresponding Baik-Rains distribution tail \cite{BR} for the stationary initial condition,
\begin{equation}\label{coincides}
-\ln\mathcal{P}\left(H,T\right)\simeq\frac{4\sqrt{2}\,\nu|H|^{3/2}}{3D|\lambda|^{1/2}T^{1/2}}.
\end{equation}
In view of the remarkable robustness of the $\lambda H \gg 1$ tail, observed for all previously studied initial conditions \citep{MKV,KMSparabola,Janas2016,MeersonSchmidt2017},
we expect Eq.~(\ref{coincides}) to hold  for arbitrary times. It would be interesting to check this prediction by extracting the $\lambda H \gg 1$  tail, at short and long times, from the exact results of Ref.~\citep{LeDoussal2017combined}.

Finally, our order parameter $\Delta$ from Eq.~(\ref{eq:Delta_def0}) can be useful for the characterization of atypical initial conditions which contribute to large deviations of different quantities in other non-equilibrium models with random initial conditions. An important example is the asymmetric exclusion process \cite{Liggett,Schuetz,Schuetz2013},
where one is interested in atypical statistics of particle current through a bond.  The discrete-lattice version of the order parameter~(\ref{eq:Delta_def0}) is the difference between the sums $\sum_{i}\left[h_{i+1}\left(t=0\right)-h_{i}\left(t=0\right)\right]^{2}$, evaluated on the two halves of the system.

\section*{ACKNOWLEDGMENTS}

We thank Pierre Le Doussal for a useful discussion. N.R.S. was supported by the Clore foundation.
A.K. was supported by NSF grant DMR-1608238.
%A.K. was supported by NSF grant ......... %DMR1306734.
N.R.S. and B.M. were supported by the Israel Science Foundation (grant No. 807/16).

\appendix

\section*{Appendix: Creating the initial condition $h\left(x,t=0\right)$}
\renewcommand{\theequation}{A\arabic{equation}}
\setcounter{equation}{0}

Here we briefly outline the optimal interface history $h\left(x,-\infty<t<0\right)$, which leads to a specified profile $h_{0}\left(x\right)$.

For stochastic dynamics in equilibrium the optimal interface history (the activation history) would coincide with the time-reversed relaxation history  \citep{Onsager}. The KPZ interface, however, is out of equlibrium even when it is in its steady state.  In order to find the activation history for a stationary KPZ interface in 1+1 dimension, one must solve the OFM equations~(\ref{eqh}) and~(\ref{eqrho}) under the conditions $h\left(x,t=0\right)=h_{0}\left(x\right)$ and $h\left(x,t\to-\infty\right)\to\text{const}$.
It is crucial that the solution lies on the invariant manifold
\begin{equation}
\label{eq:manifold}
\rho\left(x,t\right)+2\partial_{x}^{2}h\left(x,t\right)=0
\end{equation}
of the OFM equations \citep{MeersonSchmidt2017, SmithMeerson2018}. The manifold~(\ref{eq:manifold}) is related to the deterministic invariant manifold $\rho = 0$ through the transformation~(\ref{newsymmetry}).
Plugging Eq.~(\ref{eq:manifold}) into Eq.~(\ref{eqh}) leads to the equation
\begin{equation}
\label{eq:antiKPZ}
\partial_{t}h=-\partial_{x}^{2}h-\frac{1}{2}\left(\partial_{x}h\right)^{2}.
\end{equation}
Equation~(\ref{eq:antiKPZ}) does \emph{not} coincide with the time-reversed deterministic KPZ equation, due to the sign of the nonlinear term. Still, Eq.~(\ref{eq:antiKPZ}) can be solved using the Hopf-Cole transformation. Plugging $Q \equiv e^{h/2}$ into Eq.~(\ref{eq:antiKPZ}) yields the anti-diffusion equation
\begin{equation}
\partial_{t}Q=-\partial_{x}^{2}Q
\end{equation}
which can be solved backwards in time with the ``initial" condition
\begin{equation}
Q\left(x,t=0\right)=e^{h_{0}\left(x\right)/2}.
\end{equation}
Notably,  the sign in the exponent $e^{h/2}$ is opposite to that of the Hopf-Cole transformation applied to the deterministic KPZ equation.

The optimal history in terms of $h\left(x,t\right)$ is given by $h\left(x,t\right)=2\ln Q\left(x,t\right)$. The dynamical action (\ref{actiongeneral}), evaluated on $h\left(x,t\right)$ yields the interfacial cost of $h_{0}\left(x\right)$, described by Eq.~(\ref{cost}) \citep{SmithMeerson2018}. Of course, this fact makes the ``prehistoric" dynamical calculations unnecessary for the purpose of evaluating the probability of creation of $h_{0}\left(x\right)$, in analogy to what happens in equilibrium systems.

The formal condition for
the applicability condition of the OFM in the ``prehistoric" calculation is, as usual,  a large action. Let the desired height profile $h_{0}\left(x\right)$ have a characteristic height $H_0$ and width $L_0$, in the physical units. Then the OFM is applicable if
$DL_{0}/\left(\nu H_{0}^{2}\right)\ll1$. This condition is very different from the condition $\epsilon = D\lambda^2 \sqrt{T}/\nu^{5/2}\ll 1$. The latter is sufficient for the applicability of the OFM in the description  of the complete one-point height statistics at a specified time $t=T$, dealt with in the main text.


\bigskip\bigskip



\begin{thebibliography}{99}


\bibitem{KPZ}  M. Kardar, G. Parisi, and Y.-C. Zhang, Phys. Rev. Lett. \textbf{56}, 889 (1986).

\bibitem{HHZ} T. Halpin-Healy and Y.-C. Zhang, Phys. Reports \textbf{254}, 215 (1995); T. Halpin-Healy and K. A. Takeuchi,
J. Stat. Phys. \textbf{160}, 794 (2015).

\bibitem{Barabasi} A.-L. Barabasi and H. E. Stanley, {\it Fractal Concepts in Surface Growth} (Cambridge
University Press, Cambridge, UK, 1995).

\bibitem{Krug}
J. Krug, Adv. Phys. \textbf{46}, 139 (1997).



\bibitem{QS}
J. Quastel and  H. Spohn, J. Stat. Phys. \textbf{160}, 965 (2015).

\bibitem{S2016}
H. Spohn, in ``Stochastic Processes and Random Matrices", Lecture Notes of the Les Houches Summer School, vol. 104, edited by Gr\'{e}gory Schehr, Alexander Altland, Yan V. Fyodorov, and Leticia F. Cugliandolo (Oxford University Press, Oxford, 2015); arXiv:1601.00499.


\bibitem{Takeuchi2018} K. A. Takeuchi, arXiv:1708.06060.
%\bibitem{Takeuchi2018} K. A. Takeuchi, Physica A (2018), https://doi.org/10.1016/j.physa.2018.03.009.


\bibitem{Corwin}
I. Corwin, Random Matrices: Theory Appl. \textbf{1}, 1130001 (2012).

\bibitem{DMS} P. Le Doussal, S. N. Majumdar, and G. Schehr,
 EPL \textbf{113}, 60004 (2016).

\bibitem{DMRS} P. Le Doussal, S. N. Majumdar, A. Rosso, and G. Schehr,
Phys. Rev. Lett. \textbf{117}, 070403 (2016).

\bibitem{SMP} P. V. Sasorov, B. Meerson, and S. Prolhac,
J. Stat. Mech. (2017) P063203.

\bibitem{LeDoussal2017} A. Krajenbrink and P. Le Doussal, Phys. Rev. E \textbf{96}, 020102(R) (2017).

\bibitem{Halperin} B. I. Halperin and M. Lax, Phys. Rev. \textbf{148}, 722 (1966).

\bibitem{Langer} J. Zittartz and J. S. Langer, Phys. Rev. \textbf{148}, 741 (1966).

\bibitem{Lifshitz} I. M. Lifshitz, Zh. Eksp. Teor. Fiz. \textbf{53}, 743 (1967) [Sov. Phys.
JETP \textbf{26}, 462 (1968)].
\bibitem{Lifshitz1988} I. Lifshitz, S. Gredeskul, and A. Pastur, {\it Introduction to the Theory of Disordered Systems} (Wiley, New York, 1988).


\bibitem{turb1} G. Falkovich, I. Kolokolov, V. Lebedev, and A. Migdal, Phys. Rev. E \textbf{54}, 4896 (1996).

\bibitem{turb2} G. Falkovich, K. Gaw\c{e}dzki, and M. Vergassola, Rev. Mod. Phys. \textbf{73}, 913 (2001).

\bibitem{turb3} T. Grafke, R. Grauer, and T. Sch\"{a}fer, J. Phys. A \textbf{48},  333001 (2015).


\bibitem{bertini2015} L. Bertini, A. De Sole, D. Gabrielli, G. Jona-Lasinio, and C. Landim, Rev. Mod. Phys. \textbf{87}, 593 (2015).


\bibitem{EK}  V. Elgart and A. Kamenev, Phys. Rev. E \textbf{70}, 041106 (2004).

\bibitem{MS2011} B. Meerson and P.V. Sasorov, Phys. Rev. E \textbf{83}, 011129 (2011); \textbf{84}, 030101(R) (2011).

\bibitem{Mikhailov1991} A. S. Mikhailov, J. Phys. A \textbf{24}, L757 (1991).

\bibitem{GurarieMigdal1996} V. Gurarie and A. Migdal, Phys. Rev. E \textbf{54}, 4908 (1996).

\bibitem{Fogedby1998} H.C. Fogedby, Phys. Rev. E \textbf{57}, 4943 (1998).

\bibitem{Fogedby1999} H.C. Fogedby, Phys. Rev. E \textbf{59}, 5065 (1999).

\bibitem{Nakao2003} H. Nakao and A. S. Mikhailov, Chaos \textbf{13}, 953 (2003).

\bibitem{KK2007} I. V. Kolokolov and S. E. Korshunov, Phys. Rev. B \textbf{75}, 140201(R) (2007).

\bibitem{KK2008} I. V. Kolokolov and S. E. Korshunov, Phys. Rev. B \textbf{78}, 024206 (2008).

\bibitem{KK2009} I. V. Kolokolov and S. E. Korshunov, Phys. Rev. E \textbf{80}, 031107 (2009).

\bibitem{Fogedby2009} H.C. Fogedby and W. Ren, Phys. Rev. E \textbf{80}, 041116 (2009).

\bibitem{MKV} B. Meerson, E. Katzav, and A. Vilenkin, Phys. Rev. Lett. \textbf{116}, 070601 (2016).

\bibitem{KMSparabola} A. Kamenev, B. Meerson, and P. V. Sasorov, Phys. Rev. E \textbf{94}, 032108 (2016).

\bibitem{Janas2016} M. Janas, A. Kamenev, and B. Meerson, Phys. Rev. E \textbf{94}, 032133 (2016).

\bibitem{MeersonSchmidt2017} B. Meerson and J. Schmidt, J. Stat. Mech. (2017) P103207.

\bibitem{Smith2018} N. Smith, B. Meerson, and P.V. Sasorov, J. Stat. Mech. (2018) 023202.

\bibitem{MSV_3d} B. Meerson, P. V. Sasorov, and A. Vilenkin, arXiv:1712.10186.

\bibitem{SmithMeerson2018} N. R. Smith and B. Meerson, arXiv:1803.04863.


\bibitem{pinned} Without losing generality, we pin the Brownian interface at
$x = 0$.

\bibitem{IS} T. Imamura and T. Sasamoto, Phys. Rev. Lett. \textbf{108}, 190603 (2012); J.
Stat. Phys. \textbf{150}, 908 (2013).

\bibitem{Borodinetal} A. Borodin, I. Corwin, P.L. Ferrari, and B. Vet\H{o}, Mathematical Physics, Analysis and Geometry \textbf{18}, 1 (2015).

\bibitem{footnote:displacement} The evolving KPZ surface has a systematic
component $h_s(t)$. For a delta-correlated noise $d h_s(t)/dt$ is infinite. This infinity can be regularized by introducing a finite correlation length of the noise. In this case $d h_s(t)/dt$ approaches, at long times, a non-unversal constant which depends on the noise correlation length\citep{Gueudre,Hairer,S2016}. Here we define $H(t)$ as $H(t)=h(0,t)-h_s(t)$.

\bibitem{BR} J. Baik and E.M. Rains, J. Stat. Phys. \textbf{100}, 523 (2000).



\bibitem{CorwinGhosal} I. Corwin and P. Ghosal, arXiv:1802.03273.

\bibitem{KLD2018} A. Krajenbrink and P. Le Doussal,  arXiv:1802.08618.

\bibitem{Corwinetal2018} I. Corwin, P. Ghosal, A. Krajenbrink,  P. Le Doussal, and Li-Cheng Tsai, arXiv:1803.05887.

\bibitem{Schuetz} G. Sch\"utz, \textit{Exactly Solvable Models for Many-Body Systems Far From Equilibrium,} in \textit{Phase Transitions and Critical Phenomena,} Vol. 19, eds. C. Domb and J. L. Lebowitz (Academic Press, London, 2001).

\bibitem{Derrida2007} B. Derrida, J. Stat. Mech. (2007) P07023.

\bibitem{hurtadoreview} P. I. Hurtado, C. P. Espigares, J. J. del Pozo, and P. L. Garrido,
J. Stat. Phys. \textbf{154}, 214 (2014).


\bibitem{Stanley} H. E. Stanley, \textit{Introduction to Phase Transitions and Critical Phenomena} (Oxford University Press, Oxford, 1971).

\bibitem{Baek2015} Y. Baek and Y. Kafri, J. Stat. Mech. (2015) P08026.

\bibitem{Baek2017} Y. Baek, Y. Kafri, and V. Lecomte, Phys. Rev. Lett. \textbf{118}, 030604 (2017).

\bibitem{Baek2018}  Y. Baek, Y. Kafri, and V. Lecomte, J. Phys. A: Math. Theor.
\textbf{51}, 10500 (2018).


\bibitem{EW1982} S. F. Edwards and D. R. Wilkinson, Proc. R. Soc. Lond. A \textbf{381}, 17 (1982).

\bibitem{Krug1992} J. Krug, P. Meakin, and T. Halpin-Healy, Phys. Rev. A \textbf{45}, 638
(1992).



\bibitem{footnote:signlambda} Changing $\lambda$ to $-\lambda$ is  equivalent to changing $h$ to $-h$.

\bibitem{Frey1996} E. Frey, U. C. T{\"a}uber, and T. Hwa, Phys. Rev. E \textbf{53}, 4424 (1996).

\bibitem{Canet2011} L. Canet, H. Chat\'{e}, B. Delamotte, and N. Wschebor, Phys. Rev.
E \textbf{84}, 061128 (2011); \textbf{86}, 019904(E) (2012).

\bibitem{Mathey2017} S. Mathey, E. Agoritsas, T. Kloss, V. Lecomte, and L. Canet, Phys. Rev. E \textbf{95}, 032117 (2017).



\bibitem{footnote:Delta} The ``na\"{\i}ve" order parameter,  suggested in Ref. \citep{Janas2016}, suffices for the identification of the phase transition. However, it would
not render the function $F(H,\Delta)$, defined here, the desired free-energy-like behavior.

\bibitem{Chernykh} A. I. Chernykh and M. G. Stepanov, Phys. Rev. E \textbf{64}, 026306 (2001).



\bibitem{footnote:stationary_ramp_validity} When this condition holds, the nonlinear term in Eq.~(\ref{eqh}), of order $\left(H/\ell\right)^{2}$, moves the interface down a negligible amount compared to $H$ itself, and the diffusion term spreads the corner singularities on a length scale of order unity, which is much smaller than $\left|\ell\right|$. It is therefore possible to neglect the dynamics altogether.

\bibitem{KrMe} P.L. Krapivsky and B. Meerson,  Phys. Rev. E \textbf{86} 031106 (2012).



\bibitem{BD2005}  T. Bodineau and B. Derrida, Phys. Rev. E \textbf{72}, 066110 (2005).

\bibitem{BD2006}  T. Bodineau and B. Derrida, J. Stat. Phys.  \textbf{123},  277  (2006).

\bibitem{SwallowTail} V. I. Arnold, \textit{Catastrophe Theory} (Springer, Berlin, 1986).

\bibitem{TracyWidom1994} C. A. Tracy and H. Widom, Comm. Math. Phys. \textbf{159}, 174 (1994).


\bibitem{LeDoussal2017combined} P. Le Doussal, J. Stat. Mech. (2017) P053210.

\bibitem{Liggett} T. M. Liggett, \textit{Stochastic Interacting Systems: Contact, Voter and Exclusion Processes} (Springer, Berlin, 1999).



\bibitem{Schuetz2013} V. Belitsky and  G. M. Sch\"{u}tz, J. Stat. Phys.  \textbf{152}, 93 (2013).


\bibitem{Onsager} L. Onsager and S. Machlup,  Phys. Rev. \textbf{91}, 1505 (1953).

\bibitem{Gueudre} T. Gueudr\'{e}, P. Le Doussal, A. Rosso, A. Henry, and P. Calabrese,
Phys. Rev. E \textbf{86}, 041151 (2012).

\bibitem{Hairer} M. Hairer, Annals of Math. \textbf{178}, 559 (2013).

\end{thebibliography}
\end{document}